\documentclass[letterpaper]{emulateapj}
\usepackage{lscape}

\begin{document}

\title{Metallicity Effects on Dust Properties in Starbursting Galaxies}

\shorttitle{Low-Metallicity Dust in Starbursts}

\author{
C.~W.\ Engelbracht\altaffilmark{1},
G.~H.\ Rieke\altaffilmark{1},
K.~D.\ Gordon\altaffilmark{1},
J.-D.~T.\ Smith\altaffilmark{1},
M.~W.\ Werner\altaffilmark{2},
J.\ Moustakas\altaffilmark{3},
C.~N.~A.\ Willmer\altaffilmark{1},
and
L.\ Vanzi\altaffilmark{4}
}

\altaffiltext{1}{Steward Observatory, University of Arizona, Tucson, AZ 85721}
\altaffiltext{2}{Jet Propulsion Laboratory, MC 264-767, 4800 Oak Grove Drive,
Pasadena, CA 91109}
\altaffiltext{3}{Physics Department, New York University, 4 Washington Place,
New York, NY 10003}
\altaffiltext{4}{European Southern Observatory, Alonso de Cordova 3107,
Vitacura, Santiago, Chile}

\begin{abstract}

We present infrared observations of 66 starburst galaxies over a wide range of
oxygen abundances, to measure how metallicity affects their dust properties.
The data include imaging and spectroscopy from the {\it Spitzer} Space
Telescope, supplemented by groundbased near-infrared imaging.  We confirm a
strong correlation of aromatic emission with metallicity, with a threshold at
a metallicity [$12+{\rm log(O/H)}$] $\sim$ 8.  The large scatter in both the
metallicity and radiation hardness dependence of this behavior implies that it
is not due to a single effect, but to some combination. We show that the
far-infrared color temperature of the large dust grains increases towards
lower metallicity, peaking at a metallicity of 8 before turning over.  We
compute dust masses and compare them to HI masses from the literature to
derive the gas to dust ratio, which increases by nearly 3 orders of magnitude
between solar metallicity and a metallicity of 8, below which it flattens out.
The abrupt change in aromatic emission at mid-infrared wavelengths thus
appears to be reflected in the far-infrared properties, indicating that
metallicity changes affect the composition of the full range of dust grain
sizes that dominate the infrared emission. In addition, we find that the ratio
L(8~\micron)/L(TIR), important for calibrating 24~\micron\ measurements of
high redshift galaxies, increases slightly as the metallicity decreases from
$\sim$ solar to $\sim$ 50\% of solar, and then decreases by an order of
magnitude with further decreases in metallicity.  Although the great majority
of galaxies show similar patterns of behavior as described above, there are
three exceptions, SBS~0335-052E, Haro~11, and SHOC~391. Their infrared SEDs
are dominated energetically by the mid-IR near 24~\micron\ rather than by the
$60 - 200$~\micron\ region. In addition, they have very weak near infrared
outputs and their SEDs are dominated by emission by dust at wavelengths as
short as 1.8~\micron. The latter behavior indicates that the dominant star
forming episodes in them are extremely young. The component of the ISM
responsible for the usual far infrared emission appears either to be missing,
or inefficiently heated, in these three galaxies.

\end{abstract}


\section{Introduction}
\label{sec:introduction}

One of the early discoveries of infrared astronomy was the substantial
infrared output of star forming galaxies \citep{kleinmann70}.  It quickly
became apparent that the infrared emission was ubiquitous and represented a
substantial portion of the bolometric output associated with recent star
formation, since most of the ultraviolet and optical emission of the young
stars is absorbed and re-emitted in the infrared
\citep{rieke72,rieke79,soifer87}.  Since then, this phenomenon has been
studied extensively (as summarized in many reviews, e.g.,
\citet{telesco88,moorwood96,sanders96, kennicutt98, genzel00, sauvage05,
lagache05}). To first order, this work has established that the overall
properties of the best-studied star-forming and infrared-luminous galaxies are
dependent on two parameters: age and luminosity.  Thus, the spectral energy
distributions have been fitted with templates that vary primarily with
luminosity (e.g., \citet{devriendt99,chary01,dale02}), and stellar population
synthesis models demonstrate how the characteristics of the galaxy evolve with
the age of the dominant star-forming episode (e.g., \citet{rieke93,
leitherer95, engelbracht98}).

The initial studies of infrared galaxies were largely confined to relatively
luminous, massive, and hence metal-rich examples. Indications that metal-poor
low-luminosity galaxies might behave differently were found in the extensive
mid-infrared spectroscopic survey by \citet{roche91}. With improvements in
sensitivity and sophistication of mid and far infrared instrumentation, it has
become apparent that metallicity - and the accompanying changes in the
hardness of the UV radiation field - constitutes a third critical parameter
influencing the overall infrared properties of star-forming galaxies. The most
dramatic dependence applies to the mid-IR aromatic features (sometimes termed
"PAH" features).  In high metallicity galaxies, these features are quite
similar in strength and other properties \citep{roche91,dale07}.  Spectra
obtained with the Infrared Space Observatory (ISO) showed that they are
significantly weaker in a number of low-metallicity galaxies \citep{thuan99,
madden00, galliano03, galliano05}.  Observations with {\it Spitzer} have
confirmed this trend and indicated that there is a fairly sharp transition in
the relative aromatic feature strength near a metallicity of $12 + {\rm
log(O/H)} = 8$ \citep{houck04,engelbracht05, wu06, madden06}. 

Is the change in aromatic feature characteristics a direct result of the low
metallicity, or is the fundamental effect due to the increased hardness of the
interstellar UV radiation field in low metallicity environments?  Is the
change confined to the aromatic carriers, or do other components of the
interstellar dust undergo changes also?  With the sensitivity of {\it
Spitzer}, it is possible to explore these questions. In this paper, we
describe infrared measurements of a sample of 66 star-forming galaxies over a
very wide range of oxygen abundances (which we treat as measurements of the
total metallicities), $7.1 \le 12 + {\rm log (O/H)} \le 8.85$. We find
systematic changes in the behavior of the far-infrared color temperatures and
in the dust masses near the same metallicity where the aromatic feature
strength appears to change. Thus, near this metallicity, there may be a
general change in dust properties over the full range of sizes and
compositions that contribute to the observed infrared emission. 

In addition to providing insights to the composition and behavior of the
interstellar dust, an understanding of galaxy behavior with metallicity is
necessary to interpret infrared observations of galaxies at high redshift,
where their metallicities are observed to be lower (by up to a factor of order
two) than locally \citep{liang04, mouhcine06, rupke07}.  Since our study
includes this range in the overall context of infrared behavior with
metallicity, we can compare with the observed infrared properties at high
redshift. We find evidence for an increase in L(8~\micron)/L(TIR) with
decreasing metallicity down to $\sim$ half solar. This effect would result in
overestimates of L(TIR) based on observations at 24~\micron\ of galaxies at z
$\sim$ 2, consistent with the bias reported by Rigby et al.\ (2007) in studies
of luminous galaxies near this redshift. 

Our discussion begins in \S\ref{sec:data} by describing the sample, how the
data were reduced, and how the photometric and spectroscopic measurements were
made.  We describe basic properties of the galaxies in \S\ref{sec:results},
including spectral energy distributions (SEDs) and luminosities.
\S\ref{sec:metaleffects} describes the effects of metallicity on the galaxy
infrared properties. The paper is summarized in \S\ref{sec:summary}.

\section{Sample and Data Reduction}
\label{sec:data}

The sample includes well-known starbursting or star-forming galaxies from the
literature, which cover as wide a range of metallicities as possible.  This
includes I~Zw~18, which for decades since its identification as a
low-metallicity galaxy \citep{searle72} remained the lowest metallicity
star-forming galaxy known, only having been supplanted in that role recently
by SBS~0335-052W \citep{izotov05} and DDO~68 \citep{izotov07}.  Large,
metal-rich starburst galaxies like NGC~2903 and NGC~5236 (M83) are also
included.  The metallicity range in between is more-or-less evenly sampled,
with special emphasis placed on galaxies with metallicities below 8.1
(throughout this paper, we quote metallicities as oxygen abundances in units
of $12+{\rm log(O/H)}$\footnote{The solar oxygen abundance on this scale is
8.7 \citep{allende01}.}, which we sometimes abbreviate as ``Z''), which, as
discussed in \S\ref{sec:introduction}, is a critical metallicity for the
aromatic emission in the MIR.  We also gathered as many galaxies as we could
find in the range below a metallicity of 7.6, where galaxies are most likely
to be dominated by an unevolved stellar population \citep{izotov99}.  The list
of sample galaxies and their metallicities and distances is provided in
Table~\ref{tab:sample}.  All metallicities in this paper were computed by us,
using the sources indicated in the table, to ensure they lie on the same
scale.  Our new values are similar to those in the literature, with a mean
difference of 0 and a DISPERSIon of 0.1 dex, and our new values eliminated a
few significant outliers: the largest difference, Tol~2138-405, is 0.41 dex
higher than the literature value.  Any new distances quoted in this paper have
been computed using the redshift from the NASA Extragalactic Database (NED),
the Hubble flow model of \citet{mould00}, and H$_0=70$~km/s/Mpc.  Left out of
the sample were galaxies with well-known Active Galactic Nuclei (AGN).

For each galaxy in the sample, we performed imaging in the 7 photometric bands
provided by the Infrared Array Camera \citep[IRAC;][]{fazio04} and Multiband
Imaging Photometer for {\it Spitzer} \citep[MIPS;][]{rieke04} instruments on
the {\it Spitzer} Space Telescope \citep{werner04}.  We supplement these
measurements with near-infrared (NIR) imaging in the J, H, and K$_s$ bands
from the literature or new measurements using the $256\times256$
infrared camera at the Steward Observatory Bok Telescope.  The resulting suite
of imaging covers the $1-180$~\micron\ range.  For 43 of the targets, we also
obtained (or retrieved from the {\it Spitzer} archive) spectra over the
$5-40$~\micron\ range with {\it Spitzer's} Infrared Spectrograph
\citep[IRS;][]{houck04}.

\subsection{Reduction of Imaging Data}
\label{sec:imaging}

The MIPS data (24~\micron, 70~\micron, and 160~\micron) were generally
obtained using the ``photometry'' mode, in which a standard set of dithered
images are obtained (with 14, 10, and 2 images per cycle at 24~\micron,
70~\micron, and 160~\micron, respectively) resulting in a full-coverage
field-of-view of $\sim2$\arcmin.  For most sources, we obtained 2, 2, and 4
cycles of 3, 10, and 10 second images at 24~\micron, 70~\micron, and
160~\micron, respectively, while half as many cycles were obtained for the
brightest ($f_\nu(60\micron) > 10$~Jy) targets.  The one exception is the
large, nearby galaxy M83, which was observed twice in the medium-rate scan map
mode, but for which the reduction of the data was equivalent to the photometry
observations.  The data were reduced using version 3.06 of the MIPS Data
Analysis Tool \citep[DAT;][]{gordon05}, which performs all the steps (slope
fitting, calibration, and mosaicking) needed to produce an image of the
target.  We applied the techniques for removal of low-level artifacts and the
flux calibration as described by \citet{engelbracht07,gordon07,stansberry07}
at 24~\micron, 70~\micron, and 160~\micron, respectively.

The IRAC observations generally consisted of 4 dithered 30-second images in
each of 4 bands (3.6~\micron, 4.5~\micron, 5.8~\micron, and
8.0~\micron).  The data were fully reduced by the {\it Spitzer} Science Center
(SSC) pipeline.

The new NIR data presented here were obtained using the $256\times256$ camera
at the Steward Observatory Bok Telescope, over several runs between 2000 and
2006.  The typical observing sequence consists of 32, 32, and 48 dithered
exposures 30, 30, and 20 seconds in length in the J, H, and K$_s$ bands,
respectively.  The data were reduced in a standard way, by subtracting a dark
image, applying a custom flat field (made from each object's data, after
masking out the source), subtracting the background, then shifting and
averaging the data to make a single mosaic image for each target and band.
The flux calibration was determined using stars in the field that were also
measured by the 2 Micron All Sky Survey \citep[2MASS;][]{skrutskie06}.

\subsection{Photometry}
\label{sec:photometry}

Aperture photometry was performed on the imaging data using the ``imexam'' and
``imstat'' tasks in the Image Reduction and Analysis Facility
(IRAF)\footnote{IRAF is distributed by the National Optical Astronomy
Observatories, which are operated by the Association of Universities for
Research in Astronomy, Inc., under cooperative agreement with the National
Science Foundation.}.  We sized the apertures to encompass the obvious
(meaning visible above the noise level) emission from each galaxy.  We
measured the noise in each image by computing sigma for the Gaussian that best
fit the histogram of pixels in the background.  We estimated confusion noise
due to background galaxies by assuming the $5\sigma$ noise levels are 0.056,
3.2, and 45 mJy at 24, 70, and 160~\micron, respectively \citep{dole04}.  (For
this sample, this term is only significant at 160~\micron, being greater than
50\% of other noise sources for the faint and/or diffuse targets DDO~187,
HS~0822+3542, SBS~1102+606, Tol~1214-277, UGCA~292, and UM~461).  We converted
this to a per-pixel uncertainty, assuming a source beam 2 pixels in radius.
The noise appropriate to the aperture used to measure each galaxy (due both to
photon noise and confusion) was then added in quadrature to the calibration
uncertainty for each band: 2\% \citep[NIR;][]{cohen03}, 3\%
\citep[IRAC;][]{reach05}, 2\% \citep[24~\micron;][]{engelbracht07},
5\% \citep[70~\micron;][]{gordon07}, and 12\%
\citep[160~\micron;][]{stansberry07}.

The local background level was subtracted from each measurement, typically by
measuring the counts in an annular region around the source.  For galaxies
with low radial symmetry (i.e., edge-on galaxies or diffuse dwarfs without a
well-defined center), we measured the background in a rectangular region off
the galaxy, of similar size to the aperture used to measure the galaxy.  The
IRAC images in bands 1 and 2 typically contain many foreground stars.  For any
galaxy so extended ($\sim1^\prime$ or more) that an annular measurement would
not reflect accurately this stellar contribution, we measured the
``background'' (which includes a Galactic foreground) off the source, in an
aperture with a size identical to the one used to measure the source.

We applied aperture corrections to the IRAC and MIPS measurements to make them
reflect the global flux density of each target.  The aperture corrections were
generally 10\% or less, except for the correction to surface brightness in the
IRAC 5.8~\micron\ and 8.0~\micron\ bands \citep[25\%;][]{reach05} and in the
smallest apertures for the MIPS measurements, where the corrections were as
large as 17\%, 51\%, and 82\% at 24~\micron, 70~\micron, and 160~\micron,
respectively.  As the color corrections for red sources (in the {\it Spitzer}
bands) approach 10\%, we also applied color corrections to the IRAC and MIPS
measurements, which we determined by computing the power-law index that fit
the data for the band in question and the next longest band (except at
160~\micron, where used the 70/160~\micron\ ratio) and then interpolating in
the color-correction tables supplied in each instrument handbook.  The
galaxies are well-resolved in the NIR data and have colors similar to stars,
so aperture and color corrections are small and none were applied.

We present the aperture-corrected, color-corrected photometry in
Tables~\ref{tab:spitzer-photometry} and \ref{tab:ir-photometry}, where we have
converted the NIR magnitudes to Jy using the zero points from \citet{cohen03}.

The coordinates of some of the galaxies in this sample (especially the faint,
low-metallicity ones) available in NED are often uncertain by several
arcseconds.  Furthermore, for any galaxy, the peak infrared position can
differ from that measured at optical wavelengths due to extinction.  As a
convenience for the reader, we provide coordinates measured in the shortest
wavelength (and therefore highest angular resolution) {\it Spitzer} band,
3.6~\micron, in Table~\ref{tab:coords}.  Most of the galaxies have an obvious
central peak for which we measured the centroid, but a few nearby galaxies
have no obvious infrared peak and the value we report is derived from ellipse
fitting, as indicated in the table.

\subsection{Reduction of Spectral Data}
\label{sec:spectroscopy}

The spectral data were reduced using the S14 version of the
SSC pipeline.  We used the automated extractions, which treat every target
like a point source.  This is appropriate for many of the low-metallicity
galaxies, which tend to be compact (and thus have small angular extent in this
sample).  For extended sources, the mismatch of the slit sizes results in
offsets between spectral segments extracted from different slits.  We
compensate for this by applying multiplicative corrections to all the spectral
segments, forcing them to match where they overlap (and taking the
longest-wavelength segment, LL1, as the baseline).  The typical corrections for
compact sources are small, only 2\%, but can range up to a factor 2.5 for
extended targets.

\subsection{Measurements of Spectral Features}
\label{sec:spectral-measurements}

The measurements of spectral feature were all made using PAHFIT
\citep{smith07}.  The fluxes or equivalent widths of interesting features are
listed in Table~\ref{tab:spectral-measurements}.

\section{Results}
\label{sec:results}

\subsection{Spectral Energy Distributions}
\label{sec:seds}

\begin{figure} \plotone{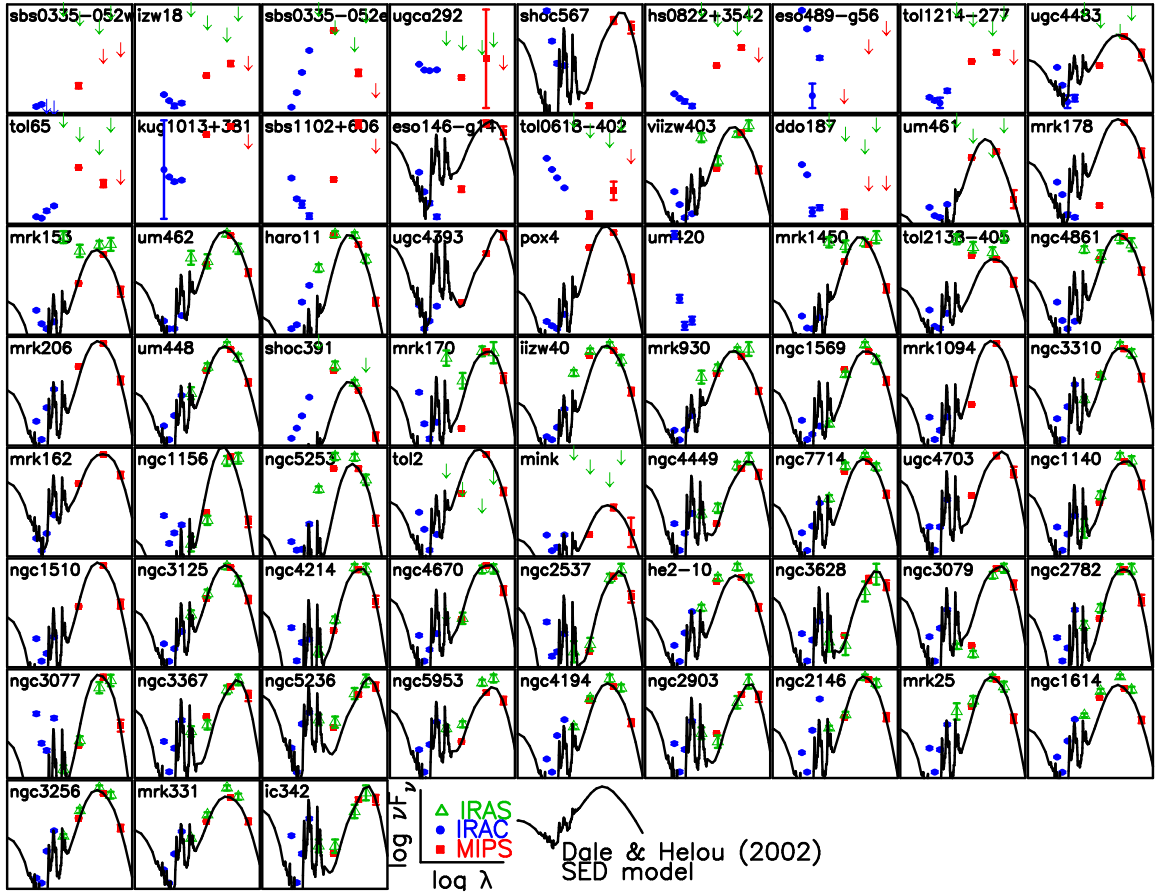} \caption{Infrared Spectral Energy
Distributions (SEDs) of starburst galaxies.  The panels are arranged in order
of increasing metallicity, left to right and top to bottom.  Where
measurements in MIPS bands 2 and 3 are available, we have chosen an SED model
as described by \citet{dale02}.\label{fig:seds}} \end{figure}

We have plotted the infrared SED of each galaxy in Figure~\ref{fig:seds}.
Where both 70~\micron\ and 160~\micron\ measurements are available, we have
fit a SED model from \citet{dale02} to the data.  The models generally fit
the high-metallicity galaxies well, but provide a less good fit at low
metallicities.  The fitting also shows that the suite of aromatic emission
features in the MIR becomes weaker at low metallicity, relative to the FIR
emission. This behavior is the primary cause of the reduction in fit quality
at low metallicity.

\begin{figure}
\plotone{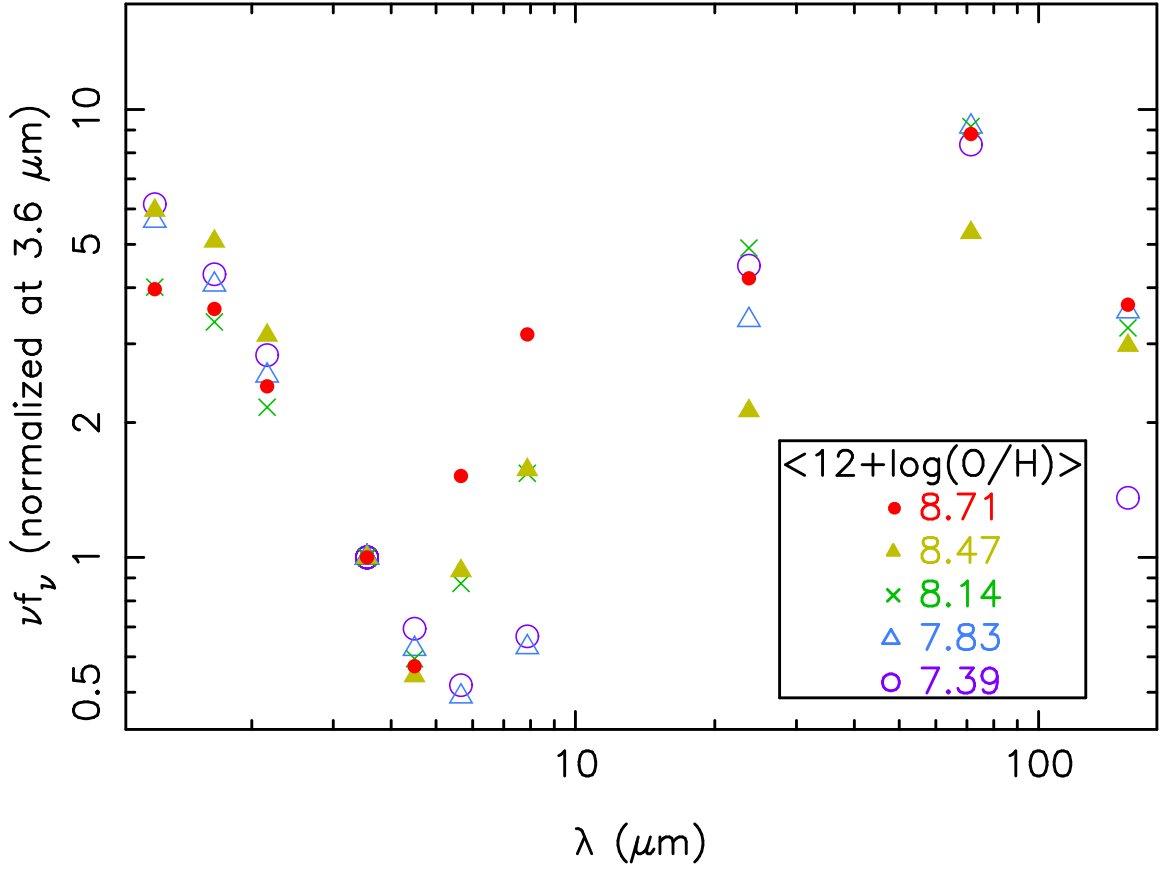}
\caption{Normalized, averaged infrared SEDs for 5 metallicity bins containing
approximately 10 galaxies each.  The SEDs of three galaxies which peak in the
MIR (SBS~0335-052E, Haro~11, and SHOC~391; see \S~\ref{sec:mirpeak}) have not
been included in any of the bins.  The inset indicates the average metallicity
of each bin.\label{fig:normalized-seds}}
\end{figure}

In fact, except in the mid-IR, the SEDs of energetically star-forming galaxies
do not vary much between 1~\micron\ and 180~\micron.  This result is
illustrated in another way in Figure~\ref{fig:normalized-seds}, where we have
plotted galaxy SEDs averaged in bins of similar metallicities and normalized
at 3.6~\micron, which traces predominantly the stellar population.  There is
little scatter between 1~\micron\ and 5~\micron\ (e.g., the K-band fluxes vary
by $\sim$15\%), where the SEDs are dominated by starlight.  This wavelength
range traces the Rayleigh-Jeans tail of most stellar SEDs and is only lightly
affected by extinction, and thus the shape of this part of the spectrum is
insensitive to the details of the stellar population.  In the FIR, the SED
shapes are also very similar, indicating the dust must be at a similar
temperature distribution.  The small scatter also indicates that the ratio of
stellar to dust luminosity does not vary substantially.  The scatter goes up
dramatically in the MIR, particularly the 5.8~\micron\ and 8.0~\micron\ bands,
which contain strong contributions from the aromatic emission features.  In
this wavelength regime, the aromatic features are systematically weak in
low-metallicity galaxies \citep{engelbracht05}, which is also reflected in the
photometry shown here.

There are only a few outstanding exceptions to this generally similar
behavior.  We discuss them in \S~\ref{sec:mirpeak} (and exclude them from
Figure~\ref{fig:normalized-seds}).

\subsection{Infrared Luminosity}
\label{sec:luminosity}

The similarity of the SEDs allows accurate determination of the total infrared
luminosities, even with sparsely sampled SEDs. 
We compare two ways to compute infrared luminosities.  One approach is to
compute the ``total infrared'' (TIR) luminosity, using MIPS measurements and
the formula given by \cite{dale02}.  This approach provides $3-1100$~\micron\
dust luminosities based on a set of SED templates.  We compare the results
using the formula to those obtained by integrating the SEDs directly.  For
each galaxy, we numerically integrate the dust emission from $3-1100$~\micron\
as follows:  We interpolate linearly the stellar-subtracted measurements (from
Table~\ref{tab:stellar-subtracted-photometry}) between 4.5~\micron\ and
160~\micron.  The dust luminosity in the shortest-wavelength IRAC band
(3.6~\micron) is both small and poorly determined (see
\S~\ref{sec:stellar-subtraction}), so we simply assume the emission goes
linearly to 0 from 4.5~\micron\ to 3~\micron.  At long wavelengths, we assume
the emission goes as a blackbody (modified by a $\lambda^{-2}$ emissivity)
between 160~\micron\ and 1100~\micron.  The results are given in
Table~\ref{tab:luminosity}, where we can see the two approaches give very
similar results.

\section{Trends with Metallicity}
\label{sec:metaleffects}

\subsection{Behavior with L(H$\alpha$)/L(TIR)}

\begin{figure}
\plotone{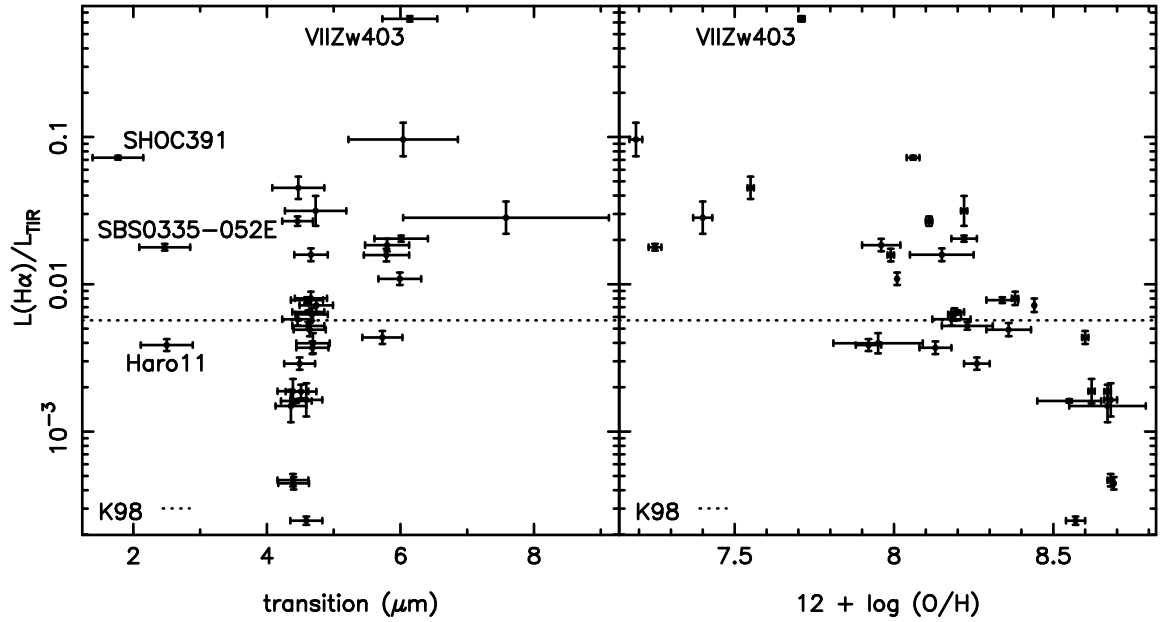}
\caption{The ratio of H$\alpha$ luminosity to TIR is plotted against the
transition wavelength (the wavelength at which the transition from stellar to
dust emission occurs) in the left panel and against metallicity in the right
panel.  The ratio predicted by the relations of \citet{kennicutt98} is marked
by a dotted line.  See \S~\ref{sec:seds} for details.  We have labelled
outliers on the plot.\label{fig:transition}}
\end{figure}

L(H$\alpha$) and L(TIR) are two star formation indicators; L(H$\alpha$)
can be taken as a measure of the relatively unobscured star formation, while 
L(TIR) is a reflection of the obscured portion (e.g., \citet{perezgonzalez06}). 
The overall similarity of the SEDs in the NIR/MIR region independent of the
relative amounts of obscured and unobscured star formation is illustrated in
Figure~\ref{fig:transition}. We plot the wavelength at which the dust
emission begins to dominate over the stellar emission, which we refer to as
the ``transition'' wavelength, against the ratio of H$\alpha$ to infrared
luminosity.  We compute the transition wavelength from the SED by fitting a
parabola to the photometric data around the local minimum in the MIR where the
Rayleigh-Jeans tail of the stellar population is dropping and the dust
emission starts to climb.  Below an H$\alpha$ to TIR ratio of 1\%, most of the
galaxies have a transition wavelength within 10\% of 4.5~\micron.  The
scatter goes up modestly at higher ratios, with a transition near 6~\micron\
for many cases. We show below that galaxies with high ratios also have
low metallicity.

Figure~\ref{fig:transition} also shows the behavior of L(H$\alpha$)/L(TIR)
with metallicity. There is a strong trend of decreasing obscuration with
decreasing metallicity. Star formation rates are often calculated using
a relation between L(TIR) and L(H$\alpha$) due to \citet{kennicutt98}, with
the tie to the star formation rate through H$\alpha$. We show this relation
as a dashed line in the figure. The assumed relation appears to be
a reasonable average. Where L(TIR) is dominant (toward high metallicity),
it can be a valid star formation indicator despite the relation in the figure,
since the unobscured star formation is a small fraction of the total. 
However, as the metallicity decreases, the relation between star formation
and L(TIR) will depart from the Kennicutt relation and L(TIR) will eventually
cease to be useful by itself to determine star formation rates (see, e.g.,
\citet{perezgonzalez06,calzetti07,kennicutt07}). 

\subsection{Stellar Photosphere Contribution to the MIR
Bands}
\label{sec:stellar-subtraction}

Global measurements of galaxies in the {\it Spitzer} bands all contain some
contribution from stellar photospheres, a contribution which must be taken
into account when measuring emission by dust.  This contribution decreases
with wavelength, both absolutely as stars are fainter and proportionally as
emission by the dust becomes brighter.  As we shall show, this contribution
varies widely from galaxy to galaxy and can be significant, especially in the
IRAC bands where it can be tens of percent.

We have measured the stellar contribution to our photometric measurements by
characterizing the stellar component in the NIR, specifically the 2 to
4~\micron\ range.  This wavelength regime strikes a balance between short
wavelengths, which can be heavily affected by extinction and details of the
star formation history and long wavelengths, which can contain significant
contributions from the dust component we are trying to measure.

We use population synthesis models to scale to longer wavelengths the stellar
emission measured at shorter wavelengths.  We have confirmed that the shapes of
the SEDs predicted by two commonly-used population synthesis models,
Starburst99 \citep{leitherer99} and PEGASE \citep{fioc97} are very similar
beyond 2~\micron\ and hardly vary with age after 5 to 15~Myr, depending on
metallicity.  Thus, our results are not sensitive to choice of age or model
within this parameter range and so we simply adopt 100~Myr Starburst99 models
as our fiducial stellar SEDs.

The wavebands available to us in the $2-4$~\micron\ range are K$_s$ and IRAC
band 1 at 3.6~\micron.  The stellar fraction computed from either of these
bands is subject to different systematic uncertainties.  IRAC band 1 will
possibly be affected by emission in the 3.3\micron\ feature, plus it is more
likely to be affected by very hot dust than the K$_s$ band.  However, the
K$_s$ band is more affected by extinction and sensitivity issues in the
available observations - it is more difficult to recover flux from faint
extended sources.  We estimate the systematic error on our scaling by
predicted stellar fluxes from both.  The stellar fluxes scaled from IRAC band
1 tend to be higher than from the K band, by an average of 45\% in our sample.
We find that we can improve the agreement by correcting for extinction, which
we measure using the extinction law of \citet{rieke85} (assuming a simple
foreground screen) and the difference between the modeled and observed
J$-$K$_s$ color.  This correction improves the agreement between the predicted
stellar fluxes, to 30\% on average.  In the absence of a clear reason to
prefer predictions from one band over the other, we adopt the average of the
3.6~\micron\ and extinction-corrected K$_s$ predictions as the stellar
fraction and use the difference between the predicted values as an indication
of the systematic uncertainty. The scale factors we used are presented in Table~\ref{tab:scale-factors},
while the stellar-subtracted fluxes are presented in
Table~\ref{tab:stellar-subtracted-photometry}.

\begin{figure}
\plotone{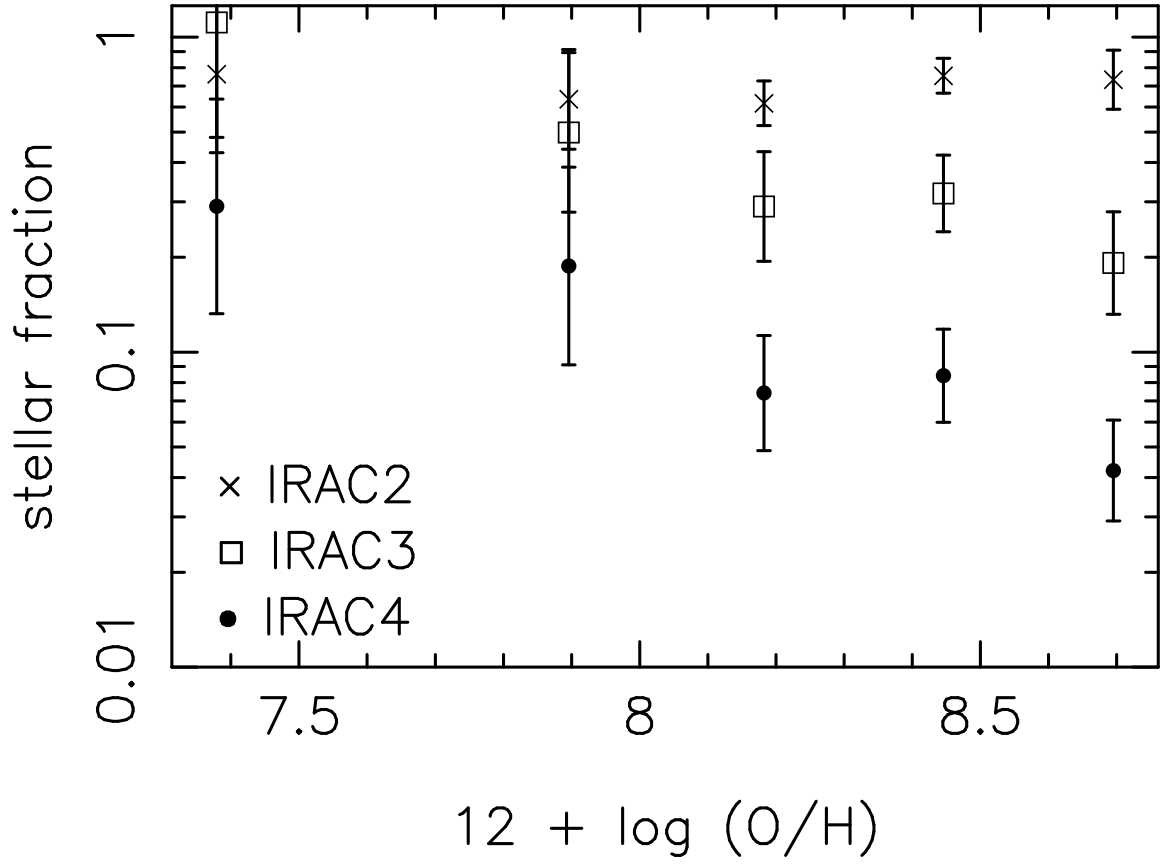}
\caption{Stellar fraction (derived from
Table~\ref{tab:stellar-subtracted-photometry}) for the three long IRAC bands,
plotted against metallicity.  The results are binned as described in
\S~\ref{sec:stellar-subtraction}.\label{fig:stellar-fraction_v_Z}}
\end{figure}

In Figure~\ref{fig:stellar-fraction_v_Z}, we plot the average stellar fraction
in the 3 longest IRAC bands (4.5~\micron, 5.8~\micron, and 8.0~\micron) in
bins of increasing metallicity.  The upper bin edges are 7.6, 8.1, 8.3, 8.6,
and 9.0, and were chosen to sample interesting physical regimes and to
maintain a statistically useful ($\sim10$) number of galaxies in each bin.
The bin center is taken to be the average metallicity of the galaxies in the
bin, and the value and uncertainty of each bin are the mean and
root-mean-square, respectively, of the stellar fractions of the galaxies in
that bin.  Two trends are most evident:  the stellar fraction decreases with
wavelength (as expected) and, at the longer wavelengths, with metallicity.
Except for a dip around a metallicity of 8, due to most of the galaxies with
unusual SEDs (see \S~\ref{sec:mirpeak}) landing in this bin, the stellar
fraction at 4.5~\micron\ is independent of Z, at a value of $~\sim70$\%, but
at 8.0~\micron, a strong trend is observed as the fraction decreases from 30\%
at the lowest metallicity to 4\% around solar metallicity.

\subsection{Dust Temperatures}
\label{sec:temperature}

We compute color temperatures for various infrared bands assuming the emission
follows a blackbody curve with an emissivity proportional to $\lambda^{-2}$.
Separate temperatures were determined for flux ratios at $24/70$~\micron,
$70/160$~\micron, and $100/160$~\micron\ by computing the temperature of the
modified blackbody curve that fit the ratio.  Uncertainties on the
temperatures reflecting the photometric uncertainties were computed via a
Monte Carlo approach.  The results are tabulated in Table~\ref{tab:tfit}.

The color temperature of the two longest bands (at 100~\micron\ and
160~\micron) should be most similar to the equilibrium temperature of the
dominant dust component, as those bands are expected to have little
contribution from stochastic heating \citep[cf.][]{popescu00,galliano05}.  We
find that the color temperature that includes 70~\micron\ data is generally
the same within the uncertainties, indicating that, for these galaxies, the
70~\micron\ band is also dominated by emission from dust in thermal
equilibrium.  We can therefore use the 70/160~\micron\ ratio to compute dust
temperatures with some physical significance.  Finally, as expected, the
24/70~\micron\ temperature is very much higher, due to the contributions at
24~\micron\ from transiently-heated grains and equilibrium emission from small
regions with very warm dust \citep[e.g.,][]{draine07}.

\begin{figure}
\plotone{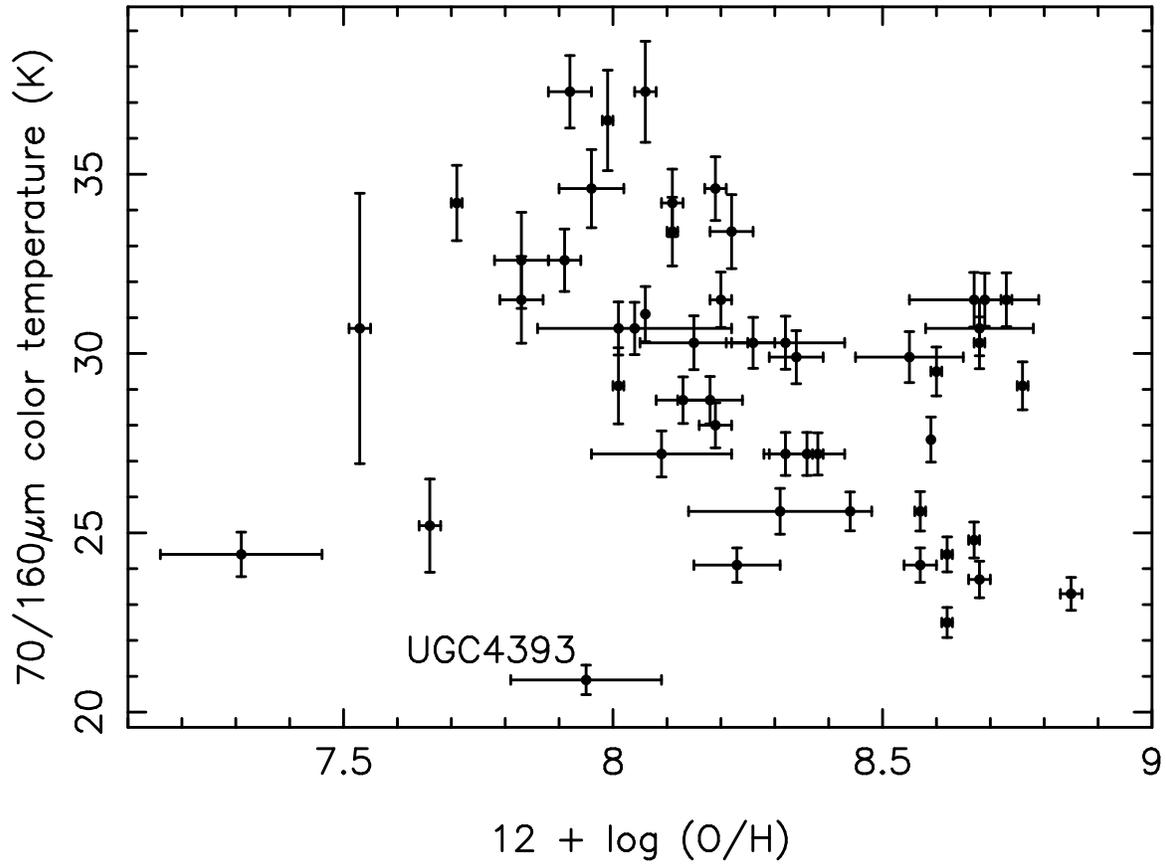}
\caption{Color temperature derived from 70~\micron\ and 160~\micron\ data as a
function of metallicity.  Temperatures and metallicities are taken from
Tables~\ref{tab:tfit} and \ref{tab:sample}, respectively.  We have labelled
outliers on the plot.
\label{fig:temperature}}
\end{figure}

We plot the 70/160~\micron\ color temperature as a function of metallicity in
Figure~\ref{fig:temperature}.  We find that, above a metallicity of $\sim8$,
dust temperature is inversely correlated with metallicity, rising from 22~K
near solar metallicity to 35~K near a metallicity of 8.  Below that
metallicity, the curve turns over, as the dust becomes cooler again.

\subsection{Dust Masses}
\label{sec:mass}

We compute dust masses using the standard formula and the
absorption coefficients from \citet{li01}.  We use the 70/160~\micron\ color
temperature from Table~\ref{tab:tfit}.  The results are presented in
Table~\ref{tab:dust-mass}.

\begin{figure}
\plotone{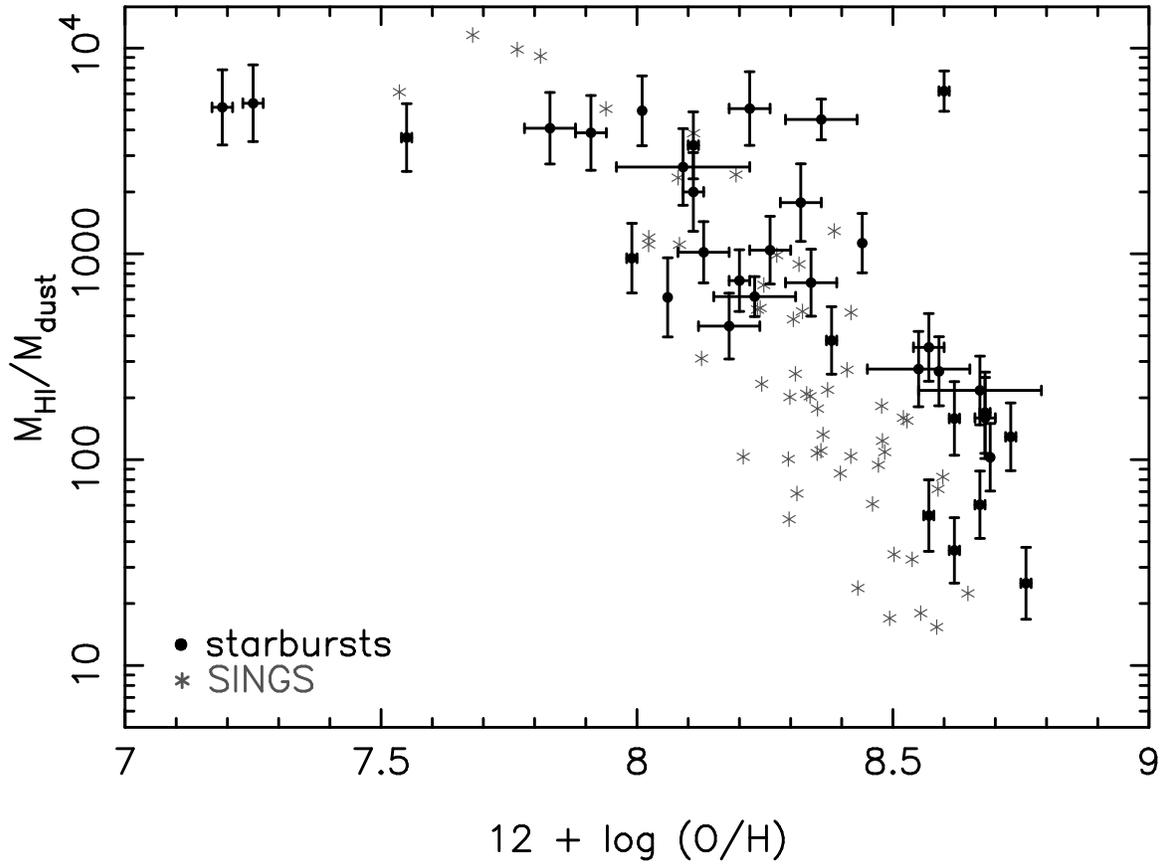}
\caption{Mass ratio of atomic gas (i.e., H$_2$ has been excluded) to dust,
plotted as a function of metallicity.  Masses and metallicities are taken from
Tables~\ref{tab:dust-mass} and \ref{tab:sample}, respectively.  The SINGS data
are taken from \citet{draine07}, where the dust mass has been computed as
described in \S\ref{sec:mass}.  \label{fig:gas-dust}}
\end{figure}

We compare the dust masses to HI masses compiled from the literature.  The HI
masses and references are summarized in Table~\ref{tab:hi}.  The ratio of ``HI
gas'' (hereafter, ``gas'') to dust mass as a function of metallicity is
plotted in Figure~\ref{fig:gas-dust}. Between solar metallicity to a
metallicity of 8, there is a steep rise, roughly as Z$^{-2.5}$, in the ratio,
or equivalently a steep decline in the dust mass.  At lower metallicity, the
gas/dust ratio flattens out.

In the same plot, we also show measurements of SINGS galaxies from
\citet{draine07}, where we have adopted their atomic gas masses\footnote{Note
that we have ignored the molecular component of the gas for this calculation,
the fraction of which is generally above average \citep[e.g.,][]{sauty03} for
the SINGS galaxies; enough to push the total gas to dust ratio over 100.} but,
to ensure we are comparing similar quantities, have recomputed the dust masses
using the prescription above.  Recomputing the dust masses makes little
difference in this plot, as our dust masses are within a factor of 2 lower, on
average, than those computed by \citet{draine07}.  At moderate (Z~$\sim8.2$)
metallicity and below, we find similar gas/dust ratios for the galaxies
discussed in this paper as for SINGS.  Above that metallicity, the SINGS
galaxies are offset below and/or to the left of the starburst galaxies in this
plot.  This is largely due to a difference in gas composition between the two
samples, since the gas in the SINGS galaxies has a higher molecular fraction
than the starbursts.  The addition of the molecular gas to the gas mass will
tend to move the SINGS galaxies closer to the starbursts in this plot.

\subsection{Aromatic Features}
\label{sec:aromatic}

Because many low metallicity galaxies are at or below the limit for high
quality IRS spectra, in \citet{engelbracht05} we introduced a photometric
approach to determining aromatic feature strength.  Here, we update that
approach and test it against the larger sample of spectra now available.  We
compute a photometric equivalent of the 7.7~\micron\ aromatic complex
(referred to hereafter as the 8~\micron\ band) using a logarithmic
interpolation of the stellar-subtracted  4.5~\micron\ and 24~\micron\ bands to
estimate the 8~\micron\ continuum, and the 8.0~\micron\ band to trace the
feature.  The equation is:

\medskip
$EW(8~\micron) = $ \\
$[f_\nu(8~\micron) - (f_\nu(4.5~\micron)^{0.66} f_\nu(24~\micron)^{0.34})]
\Delta\nu(8~\micron)$ \\
$/ (f_\nu(4.5~\micron)^{0.66}
f_\nu(24~\micron)^{0.34})(c/\lambda_{eff}(8~\micron)^2),$
\medskip

\noindent where $\Delta\nu(8~\micron)$ is the bandwidth of IRAC band 4,
$1.3\times10^{13}$~Hz, $c$ is the speed of light, and
$\lambda_{eff}(8~\micron)$ is 7.87~\micron.  

\begin{figure}
\plotone{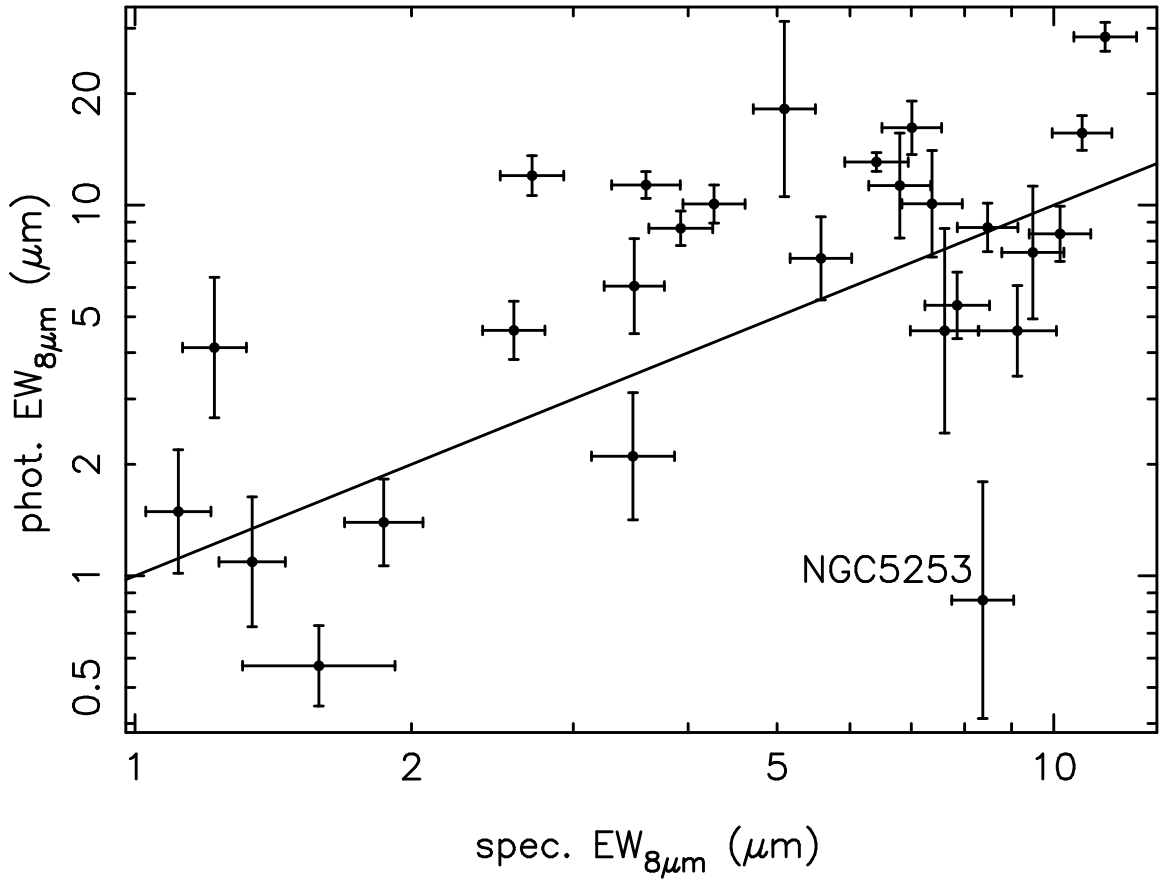}
\caption{Equivalent width of the 7.7~\micron\ aromatic complex measured
photometrically plotted against the equivalent width measured
spectroscopically (see \S~\ref{sec:aromatic}).  The error bars are propagated
from the uncertainties on the photometric and spectroscopic data points.  Not
included in the plot are galaxies known to contain an AGN (NGC~3367) or whose
strong silicate absorption renders the measurement of the 7.7~\micron\
aromatic complex highly uncertain (NGC~3079, NGC~3628, NGC~2146).  The solid
line is plotted at $phot.=spec.$ and is a reasonable fit to the data.  We have
labelled outliers on the plot.
\label{fig:phot-v-spec}}
\end{figure}

We compare the photometric measurement to the one derived directly from the
spectra (ignoring equivalent widths below 1~\micron\ as unreliable) and plot
the results in Figure~\ref{fig:phot-v-spec}.  We see that there is a
correlation between the photometric and spectroscopic measurements, confirming
the conclusion in \citet{engelbracht05} for a reduction in the relative
aromatic strength with reduced metallicity, particularly below a metallicity
of 8.  

\begin{figure}
\plotone{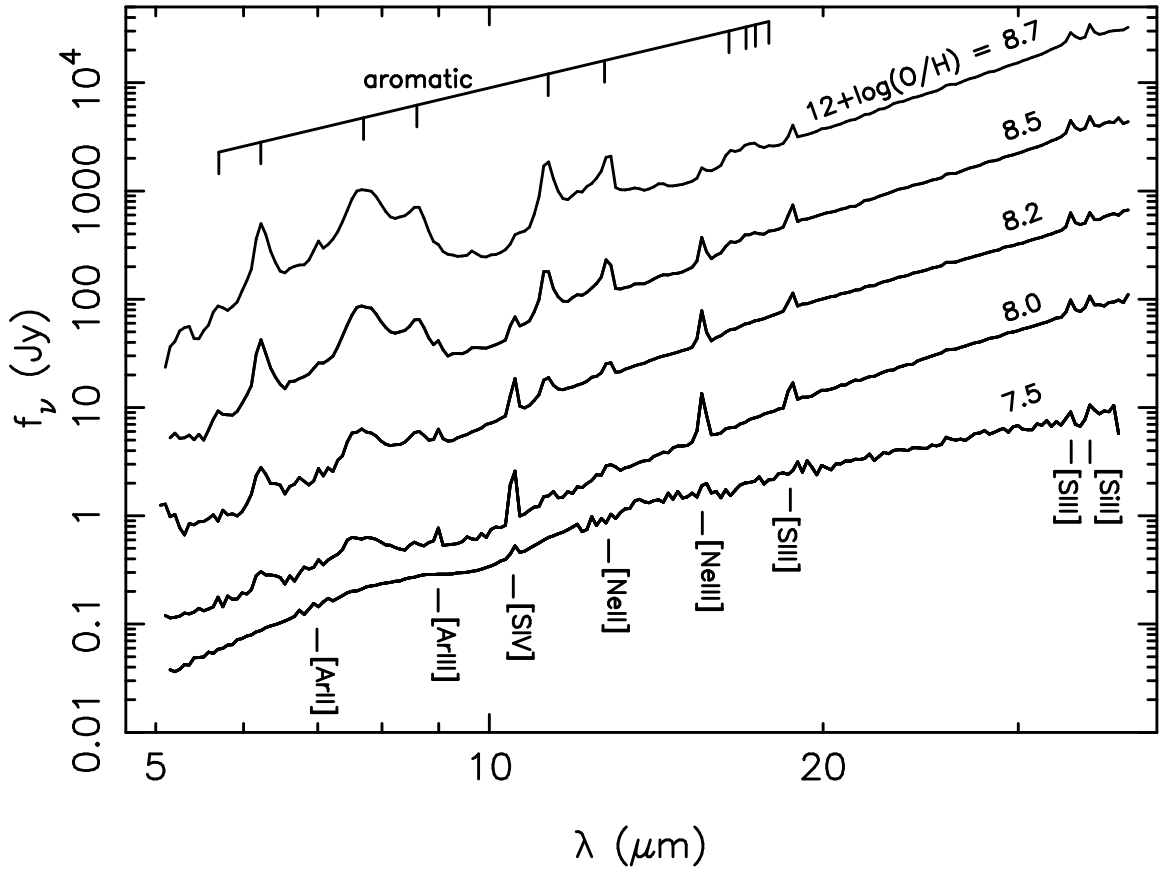}
\caption{Spectra binned according to metallicity.  From bottom to top, the
average metallicity [$12 + {\rm log (O/H)}$] is 7.5, 8.0, 8.2, 8.5, 8.7.  The spectra
were normalized at 10~\micron\ and shifted for display purposes.
\label{fig:binned-spectra}}
\end{figure}

Another way to prove the behavior of the aromatic features is to average
spectra to achieve higher signal to noise.  We have binned the spectra in
metallicity and plotted the average spectra in
Figure~\ref{fig:binned-spectra}.  The continuum slopes do not depend strongly
on metallicity, but the emission features do.  In particular, the aromatic
features steadily weaken with decreasing metallicity. However, within the
signal to noise of the averaged spectra, there is no substantial change in the
relative aromatic feature strengths - all of the transitions appear to weaken
together. Also, the ionization levels increase, most obviously traced by the
[\ion{S}{4}]$/$[\ion{S}{3}], [\ion{Ne}{3}]$/$[\ion{Ne}{2}], and
[\ion{Ar}{3}]$/$[\ion{Ar}{2}] ratios.

\begin{figure}
\plottwo{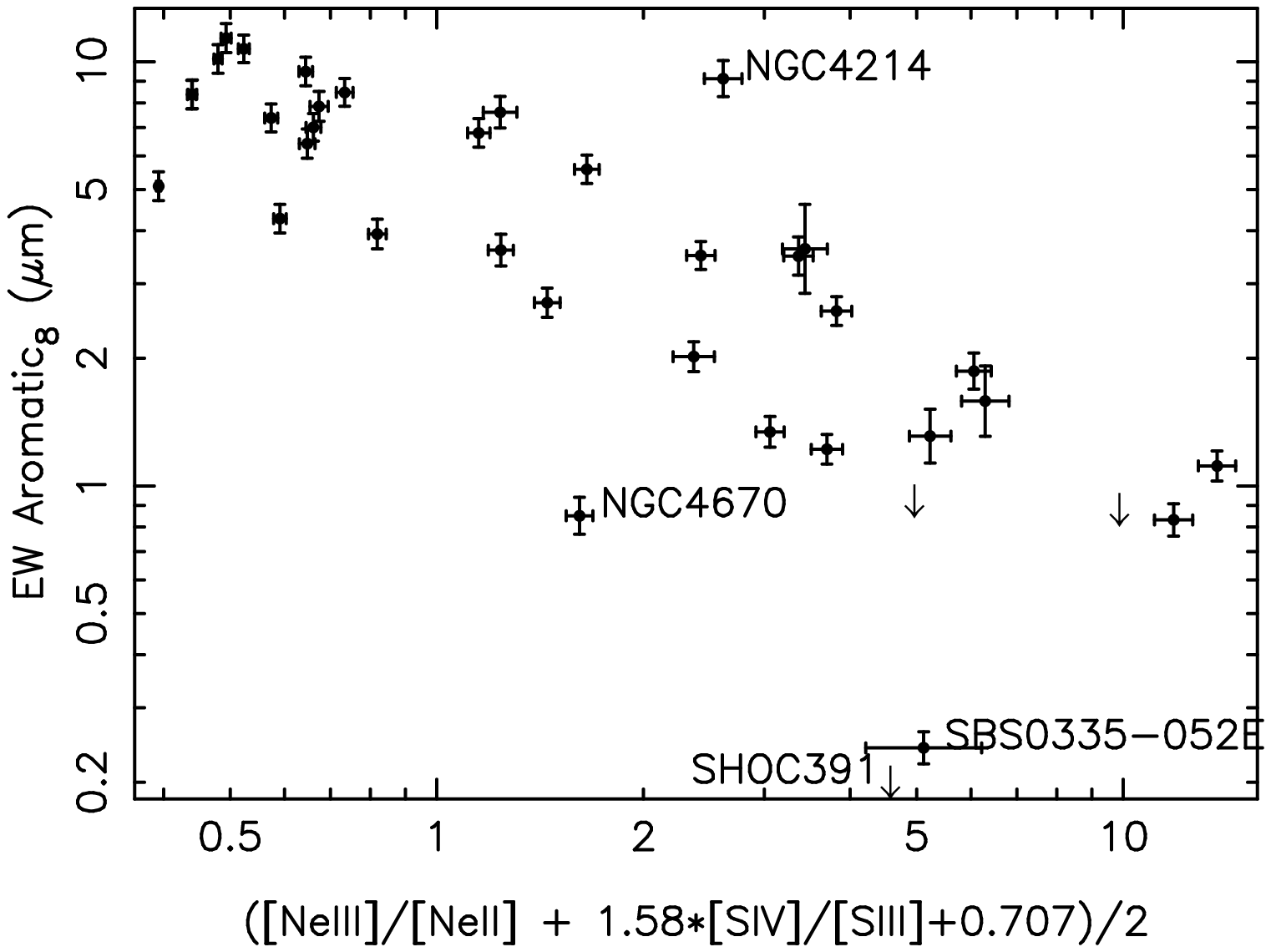}{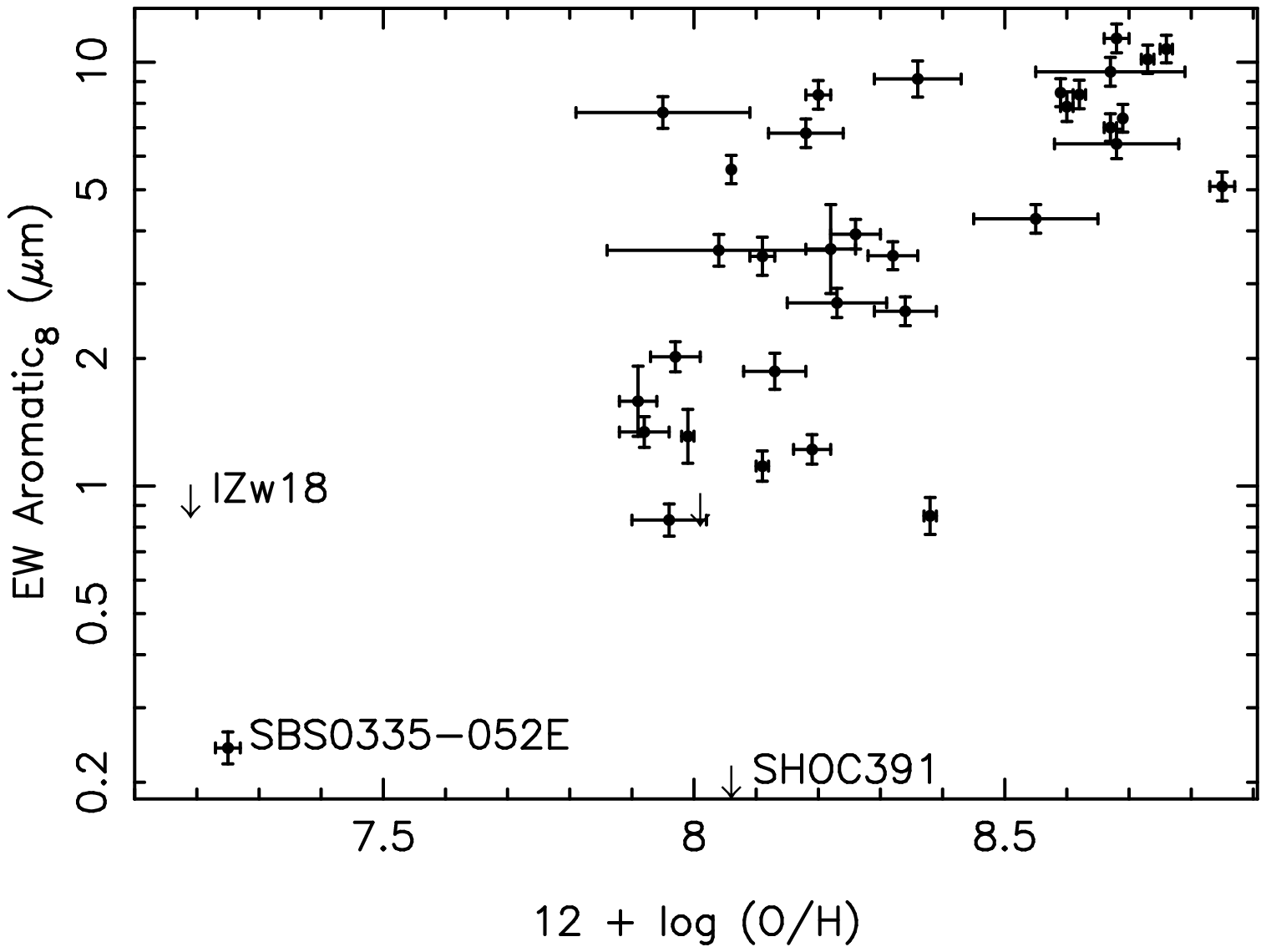}
\caption{Equivalent width of the 7.7\micron\ aromatic feature (from
Table~\ref{tab:spectral-measurements}) plotted as a function of radiation
field hardness (measured using Ne and S lines from
Table~\ref{tab:spectral-measurements}) on the left and as a function of
metallicity (from Table~\ref{tab:sample}) on the right.  Not
included in the plots are galaxies known to contain an AGN (NGC~3367) or whose
strong silicate absorption renders the measurement of the 7.7~\micron\
aromatic complex highly uncertain (NGC~3079, NGC~3628, NGC~2146).  We have
labelled outliers on the plot.
\label{fig:aromatics}}
\end{figure}

To explore what parameter affects the strength of the aromatic features, we
plot the equivalent widths (EWs) vs radiation field hardness and metallicity
(for galaxies with spectra only) in Figure~\ref{fig:aromatics}. We
characterize the radiation field with a radiation hardness index, RHI, which
is a combination of the [\ion{Ne}{3}]$/$[\ion{Ne}{2}] and
[\ion{S}{4}]$/$[\ion{S}{3}] ratios and provides a more sensitive indicator
of the hardness of the radiation field than either ratio alone. Because the
radiation hardness is judged from the IRS spectra, the measurements refer to
similar regions in the galaxies (with, for example, only weak dependence on
extinction).  There are global trends in both plots that indicate a dependence
of aromatic feature strength on both parameters.  There is nonetheless
substantial scatter (of an order of magnitude) in feature EW for a given
radiation field. At the same time, there is substantial scatter (of an order
of magnitude also) in the feature EW between metallicities of 7.9 and 8.4. The
large level of scatter suggests that neither parameter alone controls the
feature strength. This issue will be discussed further by Gordon et al.\ (2008,
in preparation).

\subsection{Behavior of L(8~\micron)/L(TIR)}

{\it Spitzer} measurements of infrared excesses in high-redshift galaxies are
often reported for only the 24~\micron\ band, because of its small beam (and
hence low level of confusion noise) and high sensitivity. Between ${\rm
z}\sim1.6-2.3$, the aromatic features in the 8~\micron\ region lie in the MIPS
24~\micron\ band and enhance the detection of star forming galaxies.  However,
over this redshift range the star formation rates are then based on deriving
L(TIR) from L(8~\micron), so any systematic change in the relationship in
these quantities from templates based on local galaxies will have implications
for the calibration of the 24~\micron\ data in terms of star formation.  Rigby
et~al.\ (2007, in prep.) summarize indications for a systematic shift in
L(8~\micron)/L(TIR) for ULIRGs at z $\sim$ 2, away from values typical of
local ULIRGs and toward the values typical of lower-luminosity local galaxies.
They suggest this change may be due to lower metallicity in the high-z ULIRGs,
which implies a substantial reduction in the amount of dust (see above) and
hence a reduction in the optical depth of the star forming regions. 

\begin{figure}
\plotone{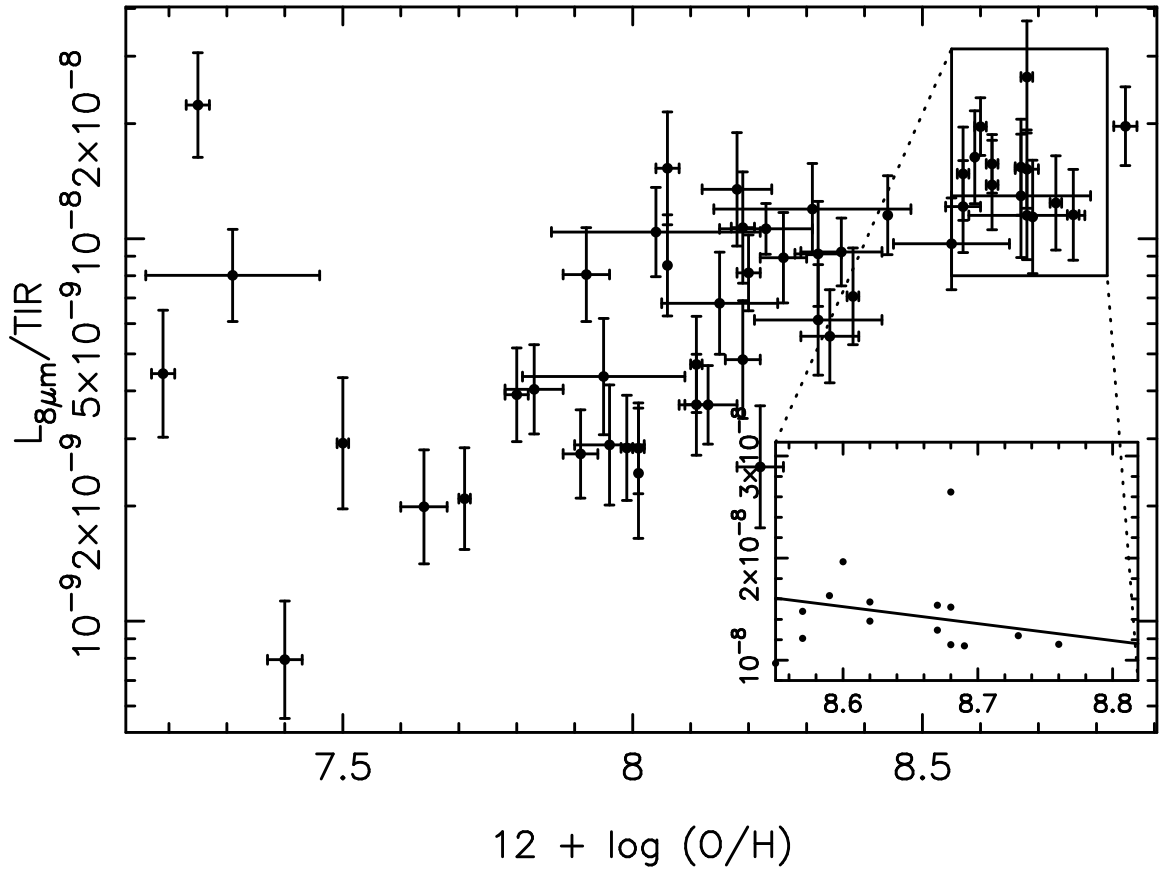}
\caption{The ratio of 8~\micron\ luminosity to total infrared luminosity (TIR;
\S~\ref{sec:luminosity}) is plotted as a function of metallicity.  The inset
plots linearly the points at high metallicity, along with a linear fit to the
data.
\label{fig:Zv8_TIR}}
\end{figure}

This suggestion can be tested with the current sample.
Figure~\ref{fig:Zv8_TIR} shows the ratio L(8~\micron) / L(TIR) as a function
of metallicity.  This ratio increases slightly from solar to $\sim1/2$ solar
metallicity and then falls toward lower metallicity, with increasing scatter.
It appears that the metallicities of luminous galaxies are reduced by a factor
of 1.5 - 2 relative to local analogs \citep{liang04, mouhcine06, rupke07}.
Therefore, the behavior in Figure~\ref{fig:Zv8_TIR} is qualitatively
consistent with the reported change at z $\sim$ 2. Making a more quantitative
comparison is not feasible because: 1.) the metallicity measurements at high
redshift are subject to substantial errors; and 2.) our sample includes few
galaxies with the large optical depths typical of local ULIRGs.

\section{Mid-Infrared Peaked SEDs}
\label{sec:mirpeak}

A major finding of this study is that the great majority of infrared-active
star-forming galaxies of widely different characteristics (morphology,
metallicity, etc.) have very similar behavior in the near, mid, and far
infrared, with the exception of the strength of their aromatic features. The
aromatic feature behavior also follows a trend controlled by the hardness of
the radiation field and probably also influenced by the metallicity (beyond
the effect of metallicity on the radiation field). In all these cases, the
bulk of the total infrared luminosity (L(TIR)) is emitted in the far infrared,
and the near infrared samples emission from a luminous population of cool,
evolved stars. 

\begin{figure}
\plotone{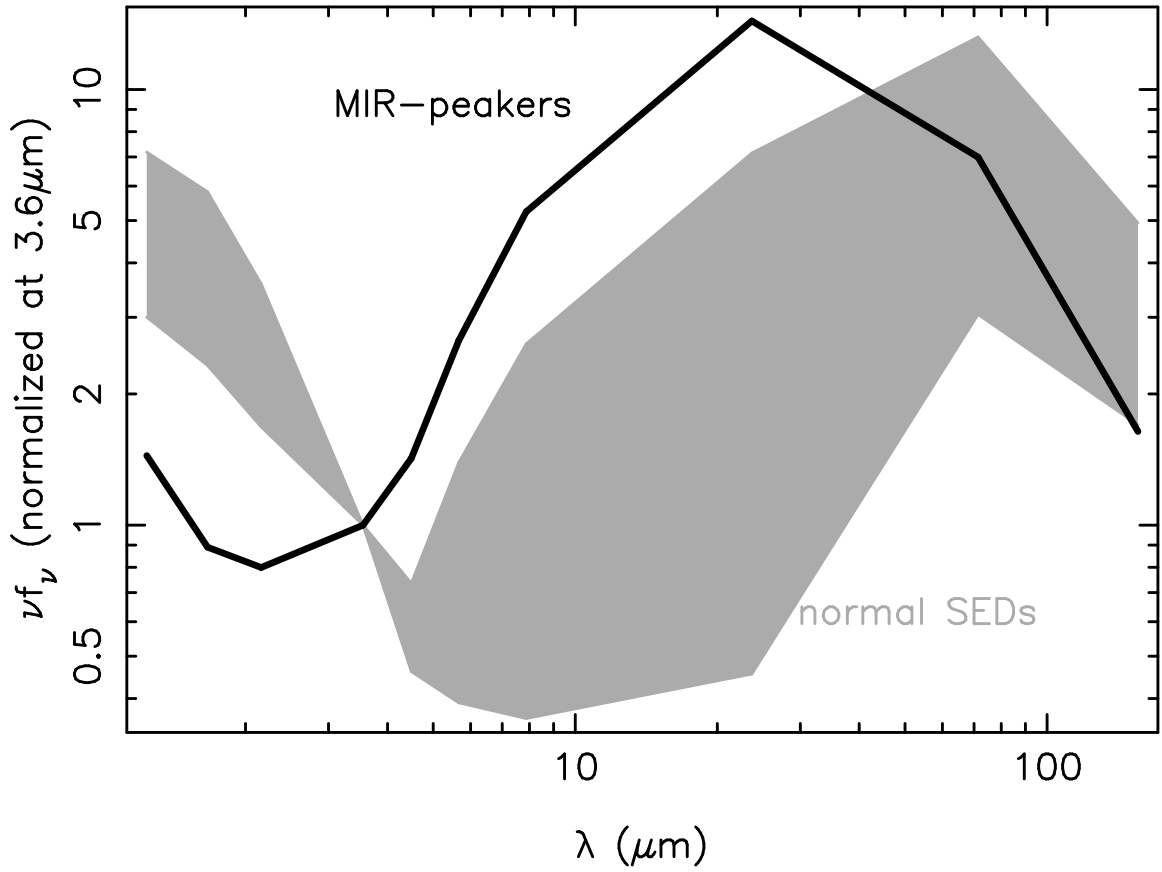}
\caption{The average SED of 3 ``MIR-peaker'' galaxies (SBS~0335-052, Haro~11,
and SHOC~391) is compared to the range of values observed in the rest of the
sample, where the grey band represents a range of $\pm1\sigma$.
\label{fig:dusty-seds}}
\end{figure}

Three galaxies (SBS 0335-052E, Haro 11, and SHOC 391) depart markedly from
these trends, with 24~\micron\ flux densities within a factor of two of those
at 70~\micron\ and very weak output in the near infrared. The latter two show
the anomalous mid-to-far infrared behavior also in the IRAS data
\citep[e.g.,][]{shupe98,schmitt06}, but it is confirmed here. The Spitzer data
make the important addition of using a sufficiently small beam at 24~\micron\
to associate the strong flux at that wavelength positively with the galaxies.
We refer to these galaxies as Mid-IR Peakers (MIRPs), for which L(TIR) is
dominated by the mid infrared.  Figure~\ref{fig:dusty-seds} compares the
average behavior of these galaxies with the range of SEDs of the other
galaxies in this sample, to emphasize the dramatic difference.  The behavior
of MIRPs  falls outside the parameter range usually explored in constructing
infrared SED templates. Their aromatic feature equivalent widths are small,
but within the trends for galaxies of similar metallicity. Similarly, their
fine structure line ratios imply relatively hot radiation fields but are not
very different from other galaxies of similar metallicity. Neither are their
optical emission line characteristics particularly distinctive (e.g., Guseva
et al.\ 2006 and references therein).  These galaxies also have relatively weak
outputs in the near infrared. The star/dust transition wavelengths in them are
near 2~\micron\ (see Figure~\ref{fig:transition}). Thus, we have determined
the stellar fraction in the J and H bands rather than K and 3.6~\micron\,
since from the SED shape these former bands are expected to be dominated by
star light while still being relatively insensitive to extinction. Any
additional uncertainties incurred by this procedure have a negligible effect
on the dust properties, since the stellar fractions in the longer wavelength
infrared bands are only a few percent at most. 

Another well-studied galaxy with some similarities is II~Zw~40, with a
relatively strong 24~\micron\ output and a weak near infrared one. This galaxy
also shows little CO first overtone absorption at 2.3~\micron, indicating
minimal contribution to its weak near infrared output from evolved stars
\citep{vanzi96}. Starburst modeling shows that the characteristics of II~Zw~40
can be explained only if it is the site of a very recent starburst, with age
$\sim$ 4  Myr. There is a very rapid growth of NIR stellar luminosity beyond
this age that quickly contradicts the observational constraints
\citep{vanzi96} Since the ratio of near infrared stellar luminosity to L(TIR)
is similar for the MIRPs as that in II~Zw~40, a similar situation must prevail
for them. This sort of situation has been found previously for other
low-luminosity, low-metallicity galaxies \citep[e.g.,][]{thompson06}. However,
it is interesting that the MIRPs have luminosity up into the LIRG range (1.4
$\times$ 10$^{11}$ L$_\odot$ for Haro 11).

However, the youth of the starbursts in these galaxies does not necessarily
explain their unique infrared SEDs. Indeed, there are a few other examples
with similarly extreme IR SEDs, but not the extreme low-luminosities in the
near infrared (Tol 65). The warm SEDs might be explained by the presence of
AGN, but there is as yet no other evidence in favor of this hypothesis. It is
difficult to take a standard model of the ISM and change the properties of the
star-forming regions in a galaxy in a way that produces such extreme SEDs (see
Dale \& Helou 2002, Fig. 3 and accompanying discussion). It seems likely
instead that the component of the ISM responsible for the very far
infrared/submm output is either largely missing, or that for some reason it is
not efficiently heated. 

\section{Summary and Conclusions}
\label{sec:summary}

We present new infrared images and spectra, both from the {\it Spitzer} Space
Telescope and from the ground, for a sample of 66 star-forming galaxies.  The
sample spans a wide range of metallicities, from the lowest known in a
star-forming galaxy to near solar.  The imaging covers the range from
1~\micron\ to 180~\micron, while the spectra cover 5~\micron\ to 40~\micron.
These observations represent the first detections of dust emission from some
of the lowest metallicity star-forming galaxies known, including the current
record holder SBS~0335-052W.

We perform photometry on the images to compute spectral energy distributions
(SEDs).  We demonstrate that, with a few exceptions, the SEDs of the galaxies
are very similar in the near infrared, where they are dominated by stellar
emission, and the far infrared, where they are dominated by emission from dust
in thermal equilibrium.  The transition from stellar to dust emission occurs
around 4.5~\micron, with little scatter for galaxies with a metallicity [$12 +
\rm {log(O/H)}$] above 8.  The scatter in this transition wavelength increases
considerably at low metallicities.  The SEDs exhibit a strong metallicity
dependence in the mid infrared, largely due to changes in the strength of the
aromatic features.  

We use a variety of simple models to derive the fraction of emission due to
stars in the mid infrared, particularly in the IRAC bands at 4.5~\micron,
5.8~\micron, and 8.0~\micron.  While this fraction is a relatively constant
70\% at 4.5~\micron, it has a strong dependence on metallicity at 8.0~\micron,
where it ranges from 4\% in metal-rich galaxies to 30\% in the
lowest-metallicity galaxies, reflecting the lower dust content and
weak aromatics in these galaxies.

We confirm previous evidence for a substantial reduction in aromatic feature
strengths below $12 + {\rm log(O/H)} \sim 8.2$, and also with increased radiation
hardness. The scatter in the aromatic EWs against both metallicity and
radiation hardness implies that this reduction is not controlled by a single
parameter, but probably by some combination of effects. 

We compute dust properties (temperature and mass) for each galaxy.  We find an
anticorrelation between dust temperature and metallicity, with equilibrium
dust temperatures of $\sim23$~K near solar metallicity up to 40~K at low
metallicity $\sim$ 8, and then falling temperatures with further reductions in
metallicity.  The derived dust masses span over 8 orders of magnitude, from
one-tenth of a solar mass to over 50~million solar masses. They exhibit a very
steep dependence on metallicity, as $\sim$ Z$^{-2.5}$ down to Z $\sim$ 8 but
have a much weaker dependence for Z $<$ 8. The change in dust behavior in
terms of aromatics, far infrared color temperature, and dust/gas mass ratio
all near Z = 8 indicates that there near this metallicity there is a general
modification of all components of the interstellar dust that dominate the
infrared emission.

We show that the ratio L(8~\micron)/L(TIR) increases with decreasing
metallicity from solar to about 50\% solar and then decreases with further
reductions in Z.  This behavior is important for interpretation of 24~\micron\
measurements of star forming galaxies at redshifts z $\sim$ 2, since the
signals for them are dominated by aromatic emission and it is likely that they
have lower metallicity than is typical of local template galaxies.

We find 3 galaxies, SBS~0335-052E, Haro~11, and SHOC~391, that have anomalous
far infrared spectral energy distributions, with weak emission near
70~\micron\ and an SED that is dominated energetically by the mid-IR near
24~\micron.  In addition, they have weak stellar outputs in the near infrared
and are dominated by dust emission down to wavelengths as short as 2~\micron.
This latter behavior indicates that they are the sites of very young dominant
star forming episodes.  Their metallicities tend to be low, but not different
from other galaxies with behavior much more similar to the majority of
infrared-active galaxies.  It appears that the dust responsible for the far
infrared emission in most galaxies is either absent, or not efficiently heated
in these three objects. They are interesting targets for further study since
they represent an extreme state of the ISM. 

\acknowledgements

This work is based in part on observations made with the {\it Spitzer Space
Telescope}, which is operated by the Jet Propulsion Laboratory, California
Institute of Technology under NASA contract 1407.  This research has made use
of the NASA/IPAC Extragalactic Database (NED) which is operated by the Jet
Propulsion Laboratory, California Institute of Technology, under contract with
the National Aeronautics and Space Administration.  Support for this work was
provided by NASA through Contract Number 1255094 issued by JPL/Caltech.

\clearpage
\LongTables
\begin{deluxetable}{lrrrrrr}
\tabletypesize{\small}
\tablecaption{Sample Galaxies and Basic Information\label{tab:sample}}
\tablewidth{0pt}
\tablehead{
\colhead{Galaxy} & 
\colhead{$12+\log\,({\rm O/H})$\tablenotemark{a}} & 
\colhead{unc.} &
\colhead{ref.} & 
\colhead{$D$ (Mpc)\tablenotemark{b}} & 
\colhead{unc.} &
\colhead{ref.}
}
\startdata
SBS 0335-052 W &  $7.10$ & $0.08$ &  1,2 &  $56$ & $17$ &  47 \\ 
I Zw 18 &  $7.19$ & $0.06$ &  3,4,5,6,7 &  $12.6$ & $3.8$ &  48 \\ 
SBS 0335-052 E &  $7.25$ & $0.05$ &  2,8 &  $57$ & $17$ &  47 \\ 
UGCA 292 &  $7.27$ & $0.08$ &  9,10 &  $3.10$ & $0.80$ &  49 \\ 
SHOC 567 &  $7.31$ & $0.16$ &  6 &  $56$ & $17$ &  47 \\ 
HS 0822+3542 &  $7.40$ & $0.06$ &  3,11,12 &  $11.0$ & $3.3$ &  50 \\ 
ESO 489-G56 &  $7.49$ & $0.06$ &  13 &  $5.00$ & $0.60$ &  49 \\ 
Tol 1214-277 &  $7.50$ & $0.05$ &  14,15,16,17 &  $120$ & $36$ &  47 \\
\enddata
\end{deluxetable}

\clearpage
\begin{landscape}
\begin{deluxetable}{llllllll}
\tabletypesize{\scriptsize}
\tablewidth{0pt}
\tablecaption{Global {\it Spitzer} photometry of sample galaxies.\label{tab:spitzer-photometry}}
\tablehead {\colhead{} & \colhead{$f_\nu(3.6 \micron)$} & \colhead{$f_\nu(4.5 \micron)$} & \colhead{$f_\nu(5.8 \micron)$} & \colhead{$f_\nu(8 \micron)$} & \colhead{$f_\nu(24 \micron)$} & \colhead{$f_\nu(70 \micron)$} & \colhead{$f_\nu(160 \micron)$} \\
\colhead{Galaxy} & \colhead{(Jy)} & \colhead{(Jy)} & \colhead{(Jy)} & \colhead{(Jy)} & \colhead{(Jy)} & \colhead{(Jy)} & \colhead{(Jy)}}
\startdata
   SBS 0335-052W & 2.34E$-$5$\pm$1.86E$-$6 & 3.47E$-$5$\pm$2.62E$-$6 & $<$3.78E$-$5 & $<$4.65E$-$5 & 6.93E$-$4$\pm$1.46E$-$4 & $<$1.75E$-$2 & $<$6.06E$-$2 \\
         I Zw 18 & 3.75E$-$4$\pm$1.24E$-$5 & 3.51E$-$4$\pm$1.24E$-$5 & 3.42E$-$4$\pm$3.83E$-$5 & 5.48E$-$4$\pm$3.74E$-$5 & 6.29E$-$3$\pm$2.27E$-$4 & 3.49E$-$2$\pm$4.79E$-$3 & $<$7.68E$-$2 \\
   SBS 0335-052E & 7.61E$-$4$\pm$2.34E$-$5 & 1.63E$-$3$\pm$4.93E$-$5 & 4.23E$-$3$\pm$1.32E$-$4 & 1.29E$-$2$\pm$3.89E$-$4 & 7.93E$-$2$\pm$1.59E$-$3 & 5.24E$-$2$\pm$6.38E$-$3 & $<$6.06E$-$2 \\
        UGCA 292 & 6.86E$-$4$\pm$2.19E$-$5 & 3.95E$-$4$\pm$1.63E$-$5 & 4.59E$-$4$\pm$5.62E$-$5 & 7.11E$-$4$\pm$7.23E$-$5 & 7.51E$-$4$\pm$1.49E$-$4 & 2.98E$-$2$\pm$2.81E+1 & $<$3.75E$-$2 \\
        SHOC 567 & 7.40E$-$3$\pm$2.22E$-$4 & 4.65E$-$3$\pm$1.40E$-$4 & 3.50E$-$3$\pm$1.26E$-$4 & 4.63E$-$3$\pm$1.43E$-$4 & 5.48E$-$3$\pm$2.92E$-$4 & 1.19E$-$1$\pm$9.83E$-$3 & 2.20E$-$1$\pm$3.22E$-$2 \\
    HS 0822+3542 & 9.50E$-$5$\pm$3.71E$-$6 & 8.89E$-$5$\pm$3.49E$-$6 & 9.17E$-$5$\pm$1.27E$-$5 & 9.36E$-$5$\pm$1.32E$-$5 & 4.02E$-$3$\pm$1.83E$-$4 & 4.05E$-$2$\pm$5.59E$-$3 & $<$5.43E$-$2 \\
     ESO 489-G56 & 1.26E$-$3$\pm$3.91E$-$5 & 8.24E$-$4$\pm$2.68E$-$5 & 9.41E$-$5$\pm$5.19E$-$5 & 5.09E$-$4$\pm$3.31E$-$5 & $<$3.84E$-$4 & $<$6.84E$-$3 & $<$2.88E$-$2 \\
    Tol 1214-277 & 7.71E$-$5$\pm$5.56E$-$6 & 7.98E$-$5$\pm$7.13E$-$6 & 8.59E$-$5$\pm$3.50E$-$5 & 2.63E$-$4$\pm$3.41E$-$5 & 5.57E$-$3$\pm$2.20E$-$4 & 3.06E$-$2$\pm$4.46E$-$3 & $<$4.74E$-$2 \\
\enddata
\tablecomments{The uncertanties listed in the table are the root-sum-square of the scatter in the counts (taken from the data), the calibration uncertainties (cf. \citet{reach05}, \citet{engelbracht07}, \citet{gordon07}, and \citet{stansberry07}), and confusion noise \citep{dole04}, while upper limits indicate galaxies that were not detected.}
\tablenotetext{a}{This object is blended with NGC~541 at 160~\micron; the error bar has been increased accordingly.}
\tablenotetext{b}{This object nearly fills the field of view at 70~\micron; based on a comparison of different background subtraction techniques, we have increased the uncertainty by 5\%.}
\tablenotetext{c}{Based on a comparison to the 24~\micron\ image, $\sim13$\% of the 70~\micron\ light falls outside the field of view; the measurement and uncertainty have been revised accordingly.}
\tablenotetext{d}{Based on a comparison to the 70~\micron\ image, $\sim9$\% of the 160~\micron\ light falls outside the field of view; the measurement and uncertainty have been revised accordingly.}
\end{deluxetable}

\clearpage
\end{landscape}
\begin{landscape}
\begin{deluxetable}{lllllllll}
\tabletypesize{\scriptsize}
\tablewidth{0pt}
\tablecaption{Global Near-Infrared and IRAS photometry of sample galaxies.\label{tab:ir-photometry}}
\tablehead {\colhead{} & \colhead{J} & \colhead{H} & \colhead{K$_s$} & \colhead{ref.} & \colhead{$f_\nu(12 \micron)$} & \colhead{$f_\nu(25 \micron)$} & \colhead{$f_\nu(60 \micron)$} & \colhead{$f_\nu(100 \micron)$} \\
\colhead{Galaxy} & \colhead{(Jy)} & \colhead{(Jy)} & \colhead{(Jy)} & \colhead{} & \colhead{(Jy)} & \colhead{(Jy)} & \colhead{(Jy)} & \colhead{(Jy)}}
\startdata
   SBS 0335-052W & 7.70E$-$5$\pm$1.03E$-$5 & 6.96E$-$5$\pm$9.36E$-$6 & 5.30E$-$5$\pm$7.13E$-$6 & 1 & $<$8.52E$-$2 & $<$9.18E$-$2 & $<$1.21E$-$1 & $<$5.67E$-$1 \\
         I Zw 18 & 7.49E$-$4$\pm$7.97E$-$5 & 6.40E$-$4$\pm$6.88E$-$5 & 5.20E$-$4$\pm$5.65E$-$5 & 2 & $<$8.52E$-$2 & $<$9.18E$-$2 & $<$1.21E$-$1 & $<$5.67E$-$1 \\
   SBS 0335-052E & 4.95E$-$4$\pm$5.20E$-$5 & 3.75E$-$4$\pm$3.96E$-$5 & 4.49E$-$4$\pm$4.73E$-$5 & 1 & $<$8.52E$-$2 & $<$9.18E$-$2 & $<$1.21E$-$1 & $<$5.67E$-$1 \\
        UGCA 292 & 2.10E$-$3$\pm$2.37E$-$4 & 2.04E$-$3$\pm$2.35E$-$4 & 1.53E$-$3$\pm$3.01E$-$4 & 3 & $<$8.52E$-$2 & $<$9.18E$-$2 & $<$1.21E$-$1 & $<$5.67E$-$1 \\
        SHOC 567 & 2.28E$-$2$\pm$2.32E$-$3 & 2.48E$-$2$\pm$2.53E$-$3 & 1.71E$-$2$\pm$1.74E$-$3 & 4 & \nodata & \nodata & \nodata & \nodata \\
    HS 0822+3542 & 1.54E$-$4$\pm$1.90E$-$5 & 3.36E$-$4$\pm$3.87E$-$5 & 1.56E$-$4$\pm$2.56E$-$5 & 5 & $<$8.52E$-$2 & $<$9.18E$-$2 & $<$1.21E$-$1 & $<$5.67E$-$1 \\
     ESO 489-G56 & 1.93E$-$3$\pm$1.97E$-$4 & 1.70E$-$3$\pm$1.75E$-$4 & 1.24E$-$3$\pm$1.28E$-$4 & 6 & \nodata & \nodata & \nodata & \nodata \\
    Tol 1214-277 & 1.73E$-$4$\pm$1.76E$-$5 & 1.28E$-$4$\pm$1.30E$-$5 & 9.91E$-$5$\pm$1.01E$-$5 & 7 & $<$8.52E$-$2 & $<$9.18E$-$2 & $<$1.21E$-$1 & $<$5.67E$-$1 \\
\enddata
\tablecomments{The uncertainties are derived from the fluctuations in the counts or from the quoted sources and are combined in quadrature with a 10\% uncertainty on the conversion to global flux \citep[cf.][]{jarrett03}, while the upper limits indicate no source was detected.}
\tablerefs{(1) \citet{vanzi00}; (2) \citet{hunt03}; (3) this paper (45"x37"
aperture); (4) this paper (26" aperture); (5) this paper (10" aperture); (6) this paper
(34"x31" aperture); (7) \citet{noeske03}; (8) this paper (45" aperture); (9)
2MASS (\citet{jarrett03}); (10) this paper (13" aperture); (11) this paper (37"x45"
aperture); (12) this paper (80"x80" aperture); (13) this paper (20" aperture); (14)
2MASS (this paper, 30" aper.); (15) \citet{cairos03}; (16) 2MASS (this paper,
15" aper.); (17) this paper (30" aperture); (18) this paper (32" aper.); (19) this paper
(29" aperture); (20) 2MASS (this paper, 20" aper.); (21) 2MASS (this paper,
15" aper.); (22) 2MASS (this paper, 40" aper.); (23) 2MASS (this paper,
49"x49" aper.).}

\end{deluxetable}

\clearpage
\end{landscape}
\begin{deluxetable}{llr}
\tablewidth{0pt}
\tablecaption{3.6\micron\ J2000.0 Coordinates\label{tab:coords}}
\tablehead {\colhead{Galaxy} & \colhead{R.A.} & \colhead{DEC.}}
\startdata
SBS 0335-052W & 03:37:38.4 & -05:02:37.0 \\
I Zw 18 & 09:34:02.1 & 55:14:27.9 \\
SBS 0335-052E & 03:37:44.1 & -05:02:40.3 \\
UGCA 292 & 12:38:39.8 & 32:45:52.0 \\
SHOC 567 & 17:19:42.1 & 61:18:51.7 \\
HS 0822+3542 & 08:25:55.5 & 35:32:31.9 \\
ESO 489-G56\tablenotemark{a} & 06:26:17.6 & -26:15:56.8 \\
Tol 1214-277 & 12:17:17.1 & -28:02:31.8 \\
\enddata
\tablenotetext{a}{Coordinates refer to the center of an ellipse fit to the outermost isotope at least $5\sigma$ above the noise level, or about 0.04 MJy/sr.}
\end{deluxetable}

\clearpage
\begin{deluxetable}{cccccc}
\tablewidth{0pt}
\tablecaption{Stellar Scale Factors, Normalized at 3.6\micron\label{tab:scale-factors}}\tablenotemark{a}
\tablehead {\colhead{Band} & \colhead{Z $< 7.70$} & \colhead{$7.70 <= \rm{Z} < 8.15$} & \colhead{$8.15 <= \rm{Z} < 8.50$} & \colhead{$8.50 <= \rm{Z} < 8.85$} & \colhead{Z $>= 8.85$}}
\startdata
J & $3.816$ & $3.143$ & $3.105$ & $3.011$ & $3.044$\\
H & $3.303$ & $3.156$ & $3.129$ & $3.179$ & $3.021$\\
K & $2.231$ & $2.215$ & $2.240$ & $2.239$ & $2.242$\\
IRAC1 & $1.000$ & $1.000$ & $1.000$ & $1.000$ & $1.000$\\
IRAC2 & $0.612$ & $0.564$ & $0.545$ & $0.524$ & $0.517$\\
IRAC3 & $0.415$ & $0.403$ & $0.398$ & $0.391$ & $0.389$\\
IRAC4 & $0.226$ & $0.227$ & $0.228$ & $0.228$ & $0.229$\\
MIPS1 & $0.032$ & $0.032$ & $0.032$ & $0.032$ & $0.032$\\
MIPS2 & $0.008$ & $0.004$ & $0.003$ & $0.003$ & $0.003$\\
MIPS3 & $0.009$ & $0.002$ & $0.001$ & $0.001$ & $0.001$\\
\enddata
\tablenotetext{a}{Computed using Starburst99, as described in \S~\ref{sec:stellar-subtraction}}
\end{deluxetable}

\clearpage
\begin{landscape}
\begin{deluxetable}{lrrrrrrr}
\tabletypesize{\scriptsize}
\tablewidth{0pt}
\tablecaption{Global stellar-subtracted {\it Spitzer} photometry of sample galaxies.\label{tab:stellar-subtracted-photometry}}
\tablehead {\colhead{} & \colhead{$f_\nu(3.6 \micron)$} & \colhead{$f_\nu(4.5 \micron)$} & \colhead{$f_\nu(5.8 \micron)$} & \colhead{$f_\nu(8 \micron)$} & \colhead{$f_\nu(24 \micron)$} & \colhead{$f_\nu(70 \micron)$} & \colhead{$f_\nu(160 \micron)$} \\
\colhead{Galaxy} & \colhead{(Jy)} & \colhead{(Jy)} & \colhead{(Jy)} & \colhead{(Jy)} & \colhead{(Jy)} & \colhead{(Jy)} & \colhead{(Jy)}}
\startdata
   SBS 0335-052W & 2.34E$-$5$\pm$1.86E$-$6 & 1.94E$-$5$\pm$1.81E$-$6 & $<$3.78E$-$5 & $<$4.65E$-$5 & 6.92E$-$4$\pm$1.46E$-$4 & $<$1.75E$-$2 & $<$6.06E$-$2 \\
         I Zw 18 & 3.75E$-$4$\pm$1.24E$-$5 & 1.58E$-$4$\pm$3.49E$-$5 & 2.09E$-$4$\pm$4.92E$-$5 & 4.71E$-$4$\pm$9.01E$-$5 & 6.28E$-$3$\pm$2.27E$-$4 & 3.49E$-$2$\pm$4.79E$-$3 & $<$7.68E$-$2 \\
   SBS 0335-052E\tablenotemark{a} & 7.61E$-$4$\pm$2.34E$-$5 & 1.50E$-$3$\pm$3.07E$-$4 & 4.14E$-$3$\pm$8.56E$-$4 & 1.28E$-$2$\pm$2.70E$-$3 & 7.93E$-$2$\pm$1.59E$-$3 & 5.24E$-$2$\pm$6.38E$-$3 & $<$6.06E$-$2 \\
        UGCA 292 & 6.86E$-$4$\pm$2.19E$-$5 & $<$1.18E$-$4 & 1.47E$-$4$\pm$1.96E$-$5 & 5.40E$-$4$\pm$6.12E$-$5 & 7.25E$-$4$\pm$1.44E$-$4 & 2.98E$-$2$\pm$2.81E+1 & $<$3.75E$-$2 \\
        SHOC 567 & 7.40E$-$3$\pm$2.22E$-$4 & $<$1.39E$-$3 & $<$9.99E$-$4 & 2.69E$-$3$\pm$1.97E$-$4 & 5.20E$-$3$\pm$2.78E$-$4 & 1.19E$-$1$\pm$9.83E$-$3 & 2.20E$-$1$\pm$3.22E$-$2 \\
    HS 0822+3542 & 9.50E$-$5$\pm$3.71E$-$6 & 2.93E$-$5$\pm$1.95E$-$6 & 4.95E$-$5$\pm$7.33E$-$6 & 7.02E$-$5$\pm$1.05E$-$5 & 4.02E$-$3$\pm$1.83E$-$4 & 4.05E$-$2$\pm$5.59E$-$3 & $<$5.43E$-$2 \\
     ESO 489-G56 & 1.26E$-$3$\pm$3.91E$-$5 & 2.55E$-$4$\pm$9.66E$-$5 & $<$9.03E$-$4 & 2.95E$-$4$\pm$1.14E$-$4 & $<$3.84E$-$4 & $<$6.84E$-$3 & $<$2.88E$-$2 \\
    Tol 1214-277 & 7.71E$-$5$\pm$5.56E$-$6 & 4.31E$-$5$\pm$1.28E$-$5 & 6.10E$-$5$\pm$3.00E$-$5 & 2.49E$-$4$\pm$7.45E$-$5 & 5.57E$-$3$\pm$2.20E$-$4 & 3.06E$-$2$\pm$4.46E$-$3 & $<$4.74E$-$2 \\
\enddata
\tablecomments{The uncertainties are the combination, in quadrature, of the uncertainties from Table~\ref{tab:spitzer-photometry} and the uncertainties in the stellar subtraction (cf. \S~\ref{sec:stellar-subtraction}).}
\tablenotetext{a}{Stellar flux scaled from J and H bands, as described in \S~\ref{sec:stellar-subtraction}}
\end{deluxetable}

\clearpage
\end{landscape}
\begin{deluxetable}{lllllllllll}
\tabletypesize{\scriptsize}
\tablewidth{0pt}
\tablecaption{Measurements of Spectral Features\tablenotemark{a}\label{tab:spectral-measurements}}
\tablehead {\colhead{} & \colhead{arom.(8\micron)}\tablenotemark{b} & \colhead{unc.} & \colhead{[\ion{Ne}{2}]} & \colhead{unc.} & \colhead{[\ion{Ne}{3}]} & \colhead{unc.} & \colhead{[\ion{S}{3}](18\micron)} & \colhead{unc.} & \colhead{[\ion{S}{4}]} & \colhead{unc.} \\
\colhead{Galaxy} & \colhead{(EW)} & \colhead{(EW)} & \colhead{(flux)} & \colhead{(flux)} & \colhead{(flux)} & \colhead{(flux)} & \colhead{(flux)} & \colhead{(flux)} & \colhead{(flux)} & \colhead{(flux)}}
\startdata
         I Zw 18 & 7.11E$-$1 & 3.07E$-$1 & 9.76E$-$4 & 2.61E$-$4 & 7.73E$-$3 & 9.22E$-$4 & 5.25E$-$3 & 8.51E$-$4 & 4.31E$-$3 & 4.29E$-$4\\
   SBS 0335-052E & 2.41E$-$1 & 2.20E$-$2 & 3.24E$-$3 & 3.41E$-$4 & 1.21E$-$2 & 1.62E$-$3 & 4.66E$-$3 & 1.67E$-$3 & 1.71E$-$2 & 9.52E$-$4\\
          UM 462 & 1.59E+0 & 3.33E$-$1 & 1.12E$-$2 & 1.01E$-$3 & 1.11E$-$1 & 5.69E$-$3 & 5.11E$-$2 & 2.86E$-$3 & 6.38E$-$2 & 3.38E$-$3\\
         Haro 11 & 1.34E+0 & 1.16E$-$1 & 3.29E$-$1 & 1.66E$-$2 & 1.27E+0 & 6.35E$-$2 & 5.94E$-$1 & 3.03E$-$2 & 5.82E$-$1 & 2.92E$-$2\\
        UGC 4393 & 7.61E+0 & 6.82E$-$1 & 4.78E$-$2 & 2.82E$-$3 & 4.38E$-$2 & 2.95E$-$3 & 3.05E$-$2 & 3.14E$-$3 & 1.64E$-$2 & 1.54E$-$3\\
           POX 4 & 8.31E$-$1 & 7.59E$-$2 & 9.48E$-$3 & 6.81E$-$4 & 1.62E$-$1 & 8.17E$-$3 & 5.18E$-$2 & 2.78E$-$3 & 1.93E$-$1 & 9.67E$-$3\\
          UM 420 & 2.02E+0 & 1.71E$-$1 & 2.51E$-$2 & 2.28E$-$3 & 6.38E$-$2 & 4.07E$-$3 & 5.22E$-$2 & 4.84E$-$3 & 4.91E$-$2 & 4.29E$-$3\\
        Mrk 1450 & 1.31E+0 & 2.06E$-$1 & 1.34E$-$2 & 1.04E$-$3 & 1.05E$-$1 & 5.64E$-$3 & 5.65E$-$2 & 3.37E$-$3 & 6.87E$-$2 & 4.11E$-$3\\
\enddata
\tablenotetext{a}{Equivalent widths are measured in \micron\ and fluxes are measured in units of $10^{-12}$erg/s/cm$^2$.}
\tablenotetext{b}{The 8~\micron\ aromatic feature is the sum of the 7.7, 8.3, and 8.6~\micron\ blends.}
\end{deluxetable}

\clearpage
\begin{deluxetable}{lllll}
\tablewidth{0pt}
\tablecaption{Infrared Luminosities\label{tab:luminosity}}
\tablehead {\colhead{Galaxy} & \colhead{TIR\tablenotemark{a}} & \colhead{unc.} & \colhead{Direct\tablenotemark{b}} & \colhead{unc.}\\
\colhead{} & \colhead{(L$_\sun$)} & \colhead{(L$_\sun$)} & \colhead{(L$_\sun$)} & \colhead{(L$_\sun$)}}
\startdata
SBS 0335-052W & 1.204e+08 & 9.788e+07 & \nodata & \nodata \\
I Zw 18 & 1.685e+07 & 8.754e+06 & 1.502e+07 & 6.457e+06 \\
SBS 0335-052E & 1.840e+09 & 7.857e+08 & 1.992e+09 & 8.468e+08 \\
UGCA 292 & 4.841e+05 & 2.751e+08 & 5.439e+05 & 3.619e+08 \\
SHOC 567 & 1.037e+09 & 4.481e+08 & 1.119e+09 & 4.778e+08 \\
HS 0822+3542 & 1.071e+07 & 5.325e+06 & 1.056e+07 & 4.573e+06 \\
ESO 489-G56 & 0.000e+00 & 0.000e+00 & 1.116e+05 & 2.374e+04 \\
Tol 1214-277 & 1.226e+09 & 5.939e+08 & 1.165e+09 & 5.017e+08 \\
\enddata
\tablecomments{The uncertainties are propagated from the photometric uncertainties.}
\tablenotetext{a}{$3-1100$~\micron\ luminosity determined using MIPS measurements and the prescription given by \citet{dale02}.}
\tablenotetext{b}{$3-1100$~\micron\ luminosity integrated directly as described in \S~\ref{sec:luminosity}.}
\end{deluxetable}

\clearpage
\begin{deluxetable}{llll}
\tablewidth{0pt}
\tablecaption{Literature H$\alpha$ Data\label{tab:halpha}}
\tablehead {\colhead{Galaxy} & \colhead{$log(\rm{H}\alpha)$} & \colhead{unc.} & \colhead{ref.}\\ \colhead{} & \colhead{(erg/s/cm$^2$)}}
\startdata
SBS 0335-052W & \nodata & \nodata & \nodata\\
I Zw 18 & -12.49\tablenotemark{a} & \nodata & 1\\
SBS 0335-052E & -12.49 & 0.01 & 2\\
UGCA 292 & -12.70\tablenotemark{a} & 0.01 & 3\\
SHOC 567 & \nodata & \nodata & \nodata\\
HS 0822+3542 & -13.1\tablenotemark{a} & 0.05 & 4\\
ESO 489-G56 & \nodata & \nodata & \nodata\\
Tol 1214-277 & \nodata & \nodata & \nodata\\
\enddata
\tablenotetext{a}{Includes [\ion{N}{2}]}
\tablerefs{(1) \citet{ott05}; (2) \citet{pustilnik04}; (3) \citet{vanzee00}; (4) \citet{gildepaz03}; (5) \citet{schmitt06}; (6) \citet{james04}; (7) \citet{thuan05}; (8) \citet{lehnert95}; (9) \citet{croft06}; (10) \citet{fabbiano90}; (11) \citet{kennicutt87}; (12) \citet{meurer06}; (13) \citet{hattori04}; (14) \citet{hoopes01}.}
\end{deluxetable}

\clearpage
\begin{deluxetable}{lllllll}
\tabletypesize{\scriptsize}
\tablewidth{0pt}
\tablecaption{Temperature fits.\label{tab:tfit}}
\tablehead {\colhead{Galaxy} & \colhead{T$_{24/70}$} & \colhead{unc.} & \colhead{T$_{70/160}$} & \colhead{unc.} & \colhead{T$_{100/160}$} & \colhead{unc.}\\
\colhead{} & \colhead{(K)}
 & \colhead{(K)}
 & \colhead{(K)}
 & \colhead{(K)}
 & \colhead{(K)}
 & \colhead{(K)}
}
\startdata
SBS 0335-052W & \nodata & \nodata & \nodata & \nodata & \nodata & \nodata \\
I Zw 18 & 56.5 & 0.8 & \nodata & \nodata & \nodata & \nodata \\
SBS 0335-052E & 81.1 & 1.4 & \nodata & \nodata & \nodata & \nodata \\
UGCA 292 & 44.0 & 1.9 & \nodata & \nodata & \nodata & \nodata \\
SHOC 567 & 47.1 & 0.4 & 24.4 & 0.6 & \nodata & \nodata \\
HS 0822+3542 & 52.2 & 0.7 & \nodata & \nodata & \nodata & \nodata \\
ESO 489-G56 & \nodata & \nodata & \nodata & \nodata & \nodata & \nodata \\
Tol 1214-277 & 56.5 & 0.9 & \nodata & \nodata & \nodata & \nodata \\
\enddata
\tablecomments{A wavelength-dependent emissivity, $\beta$, of -2 was used in all cases.  The uncertainties were derived by propagating the photometric uncertainties via a Monte Carlo technique (cf. \S~\ref{sec:temperature})}.
\end{deluxetable}

\clearpage
\begin{deluxetable}{lll}
\tablewidth{0pt}
\tablecaption{Dust Masses\label{tab:dust-mass}}
\tablehead {\colhead{Galaxy} & \colhead{M$_{dust}$} & \colhead{unc.}\\
\colhead{} & \colhead{(M$_\sun$)} & \colhead{(M$_\sun$)}}
\startdata
I Zw 18 & 2.0e+04 & 9.1e+03\\
SBS 0335-052E & 1.8e+05 & 8.0e+04\\
UGCA 292 & 3.4e+02 & 3.2e+05\\
SHOC 567 & 9.2e+05 & 4.5e+05\\
HS 0822+3542 & 7.4e+03 & 3.3e+03\\
Tol 1214-277 & 8.5e+05 & 3.8e+05\\
UGC 4483 & 5.1e+02 & 5.5e+02\\
Tol 65 & 1.8e+05 & 8.3e+04\\
\enddata
\tablecomments{Masses and uncertainties are derived as described in \S~\ref{sec:mass}.}
\end{deluxetable}

\clearpage
\begin{deluxetable}{llll}
\tablewidth{0pt}
\tablecaption{Literature HI Data\label{tab:hi}}
\tablehead {\colhead{Galaxy} & \colhead{HI Mass} & \colhead{unc.} & \colhead{ref.}\\ & \colhead{(M$_\sun$)}}
\startdata
SBS 0335-052W & \nodata & \nodata & \nodata\\
I Zw 18 & 1.03e8\tablenotemark{a} & 2.62e7 & \nodata\\
SBS 0335-052E & 9.71e8\tablenotemark{a} & 2.91e8 & \nodata\\
UGCA 292 & 5.09e7 & \nodata & 1\\
SHOC 567 & \nodata & \nodata & \nodata\\
HS 0822+3542 & \nodata & \nodata & \nodata\\
ESO 489-G56 & \nodata & \nodata & \nodata\\
Tol 1214-277 & \nodata & \nodata & 2\\
\enddata
\tablenotetext{a}{data from \citet{paturel03}, converted to mass according to \citet{pustilnik07}}
\tablerefs{(1) \citet{young03}; (2) \citet{pustilnik07}; (3) \citet{vanzee98}; (4) \citet{swaters02}; (5) \citet{thuan81}; (6) \citet{huchtmeier05}; (7) \citet{thuan04}; (8) \citet{casasola04}; (9) \citet{croft06}.}
\end{deluxetable}

\clearpage

\begin{thebibliography}{}

\bibitem[Allende Prieto et al.(2001)]{allende01} Allende Prieto, 
C., Lambert, D.~L., \& Asplund, M.\ 2001, \apjl, 556, L63

\bibitem[Bergvall \& {\"O}stlin(2002)]{bergvall02} Bergvall, N., 
\& {\"O}stlin, G.\ 2002, \aap, 390, 891

\bibitem[Cair{\'o}s et al.(2003)]{cairos03} Cair{\'o}s, L.~M., 
Caon, N., Papaderos, P., Noeske, K., V{\'{\i}}lchez, J.~M., Lorenzo, B.~G., 
\& Mu{\~n}oz-Tu{\~n}{\'o}n, C.\ 2003, \apj, 593, 312

\bibitem[Calzetti et al.(2007)]{calzetti07} Calzetti, D., et al.\ 2007, \apj, 666, 870

\bibitem[Casasola et al.(2004)]{casasola04} Casasola, V., Bettoni, D., \&
Gelletta, G.,\ 2004, \aap, 422, 941

\bibitem[Chary \& Elbaz (2001)]{chary01} Chary, R., \& Elbaz, D. 2001, \apj, 556, 562

\bibitem[Cohen et al.(2003)]{cohen03} Cohen, M., Wheaton, 
W.~A., \& Megeath, S.~T.\ 2003, \aj, 126, 1090

\bibitem[Croft et al.(2006)]{croft06} Croft, S., et al.\ 2006, \apj, 647, 1040

\bibitem[Dale \& Helou(2002)]{dale02} Dale, D.~A., \& Helou, 
G.\ 2002, \apj, 576, 159

\bibitem[Dale et al.(2007)]{dale07} Dale, D.~A., et al., accepted to ApJ

\bibitem[Devriendt et al.(1999)]{devriendt99} Devriendt, J. E. G., 
Guiderdoni, B., \& Sadat, R. 1999, \aa, 350, 381

\bibitem[Dole et al.(2004)]{dole04} Dole, H., et al.\ 2004, 
\apjs, 154, 93

\bibitem[Draine et al.(2007)]{draine07} Draine, B.~T., et al.\ 
2007, \apj, 663, 866

\bibitem[Ekholm et al.(2000)]{ekholm00} Ekholm, T., Lanoix, P., 
Teerikorpi, P., Fouqu{\'e}, P., \& Paturel, G.\ 2000, \aap, 355, 835

\bibitem[Engelbracht et al.(1998)]{engelbracht98} Engelbracht, C. W., Rieke, M. J., 
Rieke, G. H., Kelly, D. M., \& Achtermann, J. M. 1998, \apj, 505, 639

\bibitem[Engelbracht et al.(2005)]{engelbracht05} Engelbracht, C.~W.,
Gordon, K.~D., Rieke, G.~H., Werner, M.~W., Dale, D.~A., \& Latter, W.~B.\
2005, \apjl, 628, L29

\bibitem[Engelbracht et al.(2007)]{engelbracht07} Engelbracht, C.~W., 
et al.\ 2007, \pasp, 119, 994

\bibitem[Fabbiano et al.(1990)]{fabbiano90} Fabbiano, G., Heckman, 
T., \& Keel, W.~C.\ 1990, \apj, 355, 442

\bibitem[Fazio et al.(2004)]{fazio04} Fazio, G.~G., et al.\ 
2004, \apjs, 154, 10

\bibitem[Fioc \& Rocca-Volmerange(1997)]{fioc97} Fioc, M., \& 
Rocca-Volmerange, B.\ 1997, \aap, 326, 950

\bibitem[Fricke et al.(2001)]{fricke01} Fricke, K.~J., Izotov, 
Y.~I., Papaderos, P., Guseva, N.~G., \& Thuan, T.~X.\ 2001, \aj, 121, 169

\bibitem[Galliano et al.(2003)]{galliano03} Galliano, F., Madden, S. C., Jones, A. P.,
Wilson, C. D., Bernard, J.-P., \& Le Peintre, F. 2003, \aa, 407, 159

\bibitem[Galliano et al.(2005)]{galliano05} Galliano, F., Madden, 
S.~C., Jones, A.~P., Wilson, C.~D., \& Bernard, J.-P.\ 2005, \aap, 434, 867

\bibitem[Garnett(1992)]{garnett92} Garnett, D.~R.\ 1992, \aj, 
103, 1330

\bibitem[Genzel \& Cesarsky (2000)]{genzel00} Genzel, R., \& Cesarsky, C. J.
2000, \araa, 38, 761

\bibitem[Gil de Paz et al.(2003)]{gildepaz03} Gil de Paz, A., 
Madore, B.~F., \& Pevunova, O.\ 2003, \apjs, 147, 29

\bibitem[Gonzalez-Delgado et al.(1995)]{gonzalez95} 
Gonzalez-Delgado, R.~M., Perez, E., Diaz, A.~I., Garcia-Vargas, M.~L., 
Terlevich, E., \& Vilchez, J.~M.\ 1995, \apj, 439, 604

\bibitem[Gordon et al.(2005)]{gordon05} Gordon, K.~D., et al.\
2005, \pasp, 117, 503

\bibitem[Gordon et al.(2007)]{gordon07} Gordon, K.~D., et al.\ 
2007, \pasp, 119, 1019

\bibitem[Guseva et al.(2000)]{guseva00} Guseva, N.~G., Izotov, 
Y.~I., \& Thuan, T.~X.\ 2000, \apj, 531, 776

\bibitem[Guseva et al.(2007)]{guseva07} Guseva, N.~G., Izotov, 
Y.~I., Papaderos, P., \& Fricke, K.~J.\ 2007, \aap, 464, 885

\bibitem[Hadfield \& Crowther(2006)]{hadfield06} Hadfield, L.~J., 
\& Crowther, P.~A.\ 2006, \mnras, 368, 1822

\bibitem[Hattori et al.(2004)]{hattori04} Hattori, T., et al.\ 
2004, \aj, 127, 736

\bibitem[Hoopes et al.(2001)]{hoopes01} Hoopes, C.~G., 
Walterbos, R.~A.~M., \& Bothun, G.~D.\ 2001, \apj, 559, 878

\bibitem[Houck et~al.(2004)]{houck04} Houck, J. R., et~al. 2004a, \apjs, 154,
18

\bibitem[Huchtmeier et al.(2005)]{huchtmeier05} Huchtmeier, W.~K., Krishna,
G., \& Petrosian, A.\ 2005, \aap, 434, 887

\bibitem[Hunt et al.(2003)]{hunt03} Hunt, L.~K., Thuan, T.~X., 
\& Izotov, Y.~I.\ 2003, \apj, 588, 281

\bibitem[Izotov et al.(1994)]{izotov94} Izotov, Y.~I., Thuan, 
T.~X., \& Lipovetsky, V.~A.\ 1994, \apj, 435, 647

\bibitem[Izotov et al.(1997)]{izotov97} Izotov, Y.~I., Foltz, 
C.~B., Green, R.~F., Guseva, N.~G., \& Thuan, T.~X.\ 1997, \apjl, 487, L37

\bibitem[Izotov \& Thuan(1998)]{izotov98} Izotov, Y.~I., \& 
Thuan, T.~X.\ 1998, \apj, 500, 188

\bibitem[Izotov \& Thuan(1999)]{izotov99} Izotov, Y.~I., \& 
Thuan, T.~X.\ 1999, \apj, 511, 639

\bibitem[Izotov et al.(1999b)]{izotov99b} Izotov, Y.~I., Chaffee, 
F.~H., Foltz, C.~B., Green, R.~F., Guseva, N.~G., \& Thuan, T.~X.\ 1999, 
\apj, 527, 757

\bibitem[Izotov et al.(2001)]{izotov01} Izotov, Y.~I., Chaffee, 
F.~H., \& Green, R.~F.\ 2001, \apj, 562, 727

\bibitem[Izotov \& Thuan(2004)]{izotov04} Izotov, Y.~I., \& 
Thuan, T.~X.\ 2004, \apj, 602, 200

\bibitem[Izotov et al.(2004)]{izotov04b} Izotov, Y.~I., 
Papaderos, P., Guseva, N.~G., Fricke, K.~J., \& Thuan, T.~X.\ 2004, \aap, 
421, 539

\bibitem[Izotov et al.(2005)]{izotov05} Izotov, Y.~I., Thuan, 
T.~X., \& Guseva, N.~G.\ 2005, \apj, 632, 210

\bibitem[Izotov et al.(2006)]{izotov06} Izotov, Y.~I., 
Stasi{\'n}ska, G., Meynet, G., Guseva, N.~G., \& Thuan, T.~X.\ 2006, \aap, 
448, 955

\bibitem[Izotov \& Thuan(2007)]{izotov07} Izotov, Y.~I., \& 
Thuan, T.~X.\ 2007, \apj, 665, 1115

\bibitem[James et al.(2004)]{james04} James, P.~A., et al.\ 
2004, \aap, 414, 23

\bibitem[Jarrett et al.(2003)]{jarrett03} Jarrett, T.~H., 
Chester, T., Cutri, R., Schneider, S.~E., \& Huchra, J.~P.\ 2003, \aj, 125, 
525

\bibitem[Karachentsev et al.(2004)]{karachentsev04} Karachentsev, 
I.~D., Karachentseva, V.~E., Huchtmeier, W.~K., \& Makarov, D.~I.\ 2004, 
\aj, 127, 2031

\bibitem[Karachentsev(2005)]{karachentsev05} Karachentsev, I.~D.\ 
2005, \aj, 129, 178

\bibitem[Kennicutt et al.(1987)]{kennicutt87} Kennicutt, R.~C., 
Jr., Roettiger, K.~A., Keel, W.~C., van der Hulst, J.~M., \& Hummel, E.\ 
1987, \aj, 93, 1011

\bibitem[Kennicutt(1998)]{kennicutt98} Kennicutt, R.~C., Jr.\ 1998, 
\araa, 36, 189

\bibitem[Kennicutt et al.(2007)]{kennicutt07} Kennicutt, R.~C., 
Jr., et al.\ 2007, ArXiv e-prints, 708, arXiv:0708.0922

\bibitem[Kleinmann \& Low (1970)]{kleinmann70} Kleinmann, D. E., \& Low, F. J.
1970, \apjl, 159, L165

\bibitem[Kniazev \& Pustil'Nik(1998)]{kniazev98} Kniazev, A.~Y., 
\& Pustil'Nik, S.~A.\ 1998, Bull.~Special Astrophys.~Obs., 46, 23

\bibitem[Kniazev et al.(2000)]{kniazev00} Kniazev, A.~Y., et al.\ 
2000, \aap, 357, 101

\bibitem[Kniazev et al.(2003)]{kniazev03} Kniazev, A.~Y., Grebel, 
E.~K., Hao, L., Strauss, M.~A., Brinkmann, J., \& Fukugita, M.\ 2003, 
\apjl, 593, L73

\bibitem[Kniazev et al.(2004)]{kniazev04} Kniazev, A.~Y., 
Pustilnik, S.~A., Grebel, E.~K., Lee, H., \& Pramskij, A.~G.\ 2004, \apjs, 
153, 429

\bibitem[Kobulnicky \& Skillman(1996)]{kobulnicky96} Kobulnicky, 
H.~A., \& Skillman, E.~D.\ 1996, \apj, 471, 211

\bibitem[Kobulnicky et al.(1997a)]{kobulnicky97a} Kobulnicky, H.~A., 
Skillman, E.~D., Roy, J.-R., Walsh, J.~R., \& Rosa, M.~R.\ 1997, \apj, 477, 
679

\bibitem[Kobulnicky \& Skillman(1997b)]{kobulnicky97b} Kobulnicky, 
H.~A., \& Skillman, E.~D.\ 1997, \apj, 489, 636

\bibitem[Kobulnicky et al.(1999)]{kobulnicky99} Kobulnicky, H.~A., 
Kennicutt, R.~C., Jr., \& Pizagno, J.~L.\ 1999, \apj, 514, 544

\bibitem[Kong \& Cheng(1999)]{kong99} Kong, X., \& Cheng, 
F.~Z.\ 1999, \aap, 351, 477

\bibitem[Kong \& Cheng(2002)]{kong02} Kong, X., \& Cheng, 
F.~Z.\ 2002, \aap, 389, 845

\bibitem[Kunth \& {\"O}stlin(2000)]{kunth00} Kunth, D., \& {\"O}stlin, G.\ 2000,
\aapr, 10, 1

\bibitem[Lagache et al.(2005)]{lagache05} Lagache, G., Puget, J.-L., \& Dole,
H. 2005, \araa, 43, 727

\bibitem[Lee et al.(2003)]{lee03} Lee, H., McCall, M.~L., 
Kingsburgh, R.~L., Ross, R., \& Stevenson, C.~C.\ 2003, \aj, 125, 146

\bibitem[Lehnert \& Heckman(1995)]{lehnert95} Lehnert, M.~D., \& 
Heckman, T.~M.\ 1995, \apjs, 97, 89

\bibitem[Leitherer \& Heckman (1995)]{leitherer95} Leitherer, C. \& Heckman,
T. M. 1995, \apjs, 96, 9

\bibitem[Leitherer et al.(1999)]{leitherer99} Leitherer, C., et
al.\ 1999, \apjs, 123, 3

\bibitem[Liang et al.(2004)]{liang04} Liang, Y. C., Hammer, F., Flores, H.,
Elbaz, D., Marcillac, D., \& Cesarsky, C. J. 2004, \aa, 423, 867

\bibitem[Li \& Draine(2001)]{li01} Li, A., \& Draine, B.~T.\ 
2001, \apj, 554, 778

\bibitem[L{\'{\i}}pari et al.(2000)]{lipari00} L{\'{\i}}pari, 
S., D{\'{\i}}az, R., Taniguchi, Y., Terlevich, R., Dottori, H., \& 
Carranza, G.\ 2000, \aj, 120, 645

\bibitem[Lisenfeld \& Ferrara(1998)]{lisenfeld98} Lisenfeld, U., \& 
Ferrara, A.\ 1998, \apj, 496, 145

\bibitem[Madden(2000)]{madden00} Madden, S.~C.\ 2000, New 
Astronomy Review, 44, 249

\bibitem[Madden et al.(2006)]{madden06} Madden, S.~C., Galliano, 
F., Jones, A.~P., \& Sauvage, M.\ 2006, \aap, 446, 877

\bibitem[Masegosa et al.(1994)]{masegosa94} Masegosa, J., Moles, 
M., \& Campos-Aguilar, A.\ 1994, \apj, 420, 576

\bibitem[McCall et al.(1985)]{mccall85} McCall, M.~L., Rybski, 
P.~M., \& Shields, G.~A.\ 1985, \apjs, 57, 1

\bibitem[M{\'e}ndez et al.(1999)]{mendez99} M{\'e}ndez, D.~I., 
Cair{\'o}s, L.~M., Esteban, C., \& V{\'{\i}}lchez, J.~M.\ 1999, \aj, 117, 
1688

\bibitem[Meurer et al.(2006)]{meurer06} Meurer, G.~R., et al.\ 
2006, \apjs, 165, 307

\bibitem[Moorwood (1996)]{moorwood96} Moorwood, A. F. M. 1996, Sp. Sci. Rev.,
77, 303

\bibitem[Mouhcine et al.(2006)]{mouhcine06} Mouhcine, M., Bamford, S. P.,
Arag\'on-Salamanca, A., Nakamura, O., \& Milvang-Jensen, B. 2006, \mnras, 369,
891

\bibitem[Mould et al.(2000)]{mould00} Mould, J.~R., et al.\ 
2000, \apj, 545, 547

\bibitem[Moustakas \& Kennicutt(2006)]{moustakas06} Moustakas, J., 
\& Kennicutt, R.~C., Jr.\ 2006, \apj, 651, 155

\bibitem[Noeske et al.(2003)]{noeske03} Noeske, K.~G., 
Papaderos, P., Cair{\'o}s, L.~M., \& Fricke, K.~J.\ 2003, \aap, 410, 481

\bibitem[Nagao et al.(2006)]{nagao06} Nagao, T., Maiolino, R., 
\& Marconi, A.\ 2006, \aap, 459, 85

\bibitem[{\"O}stlin(2000)]{oestlin00} {\"O}stlin, G.\ 2000, 
\apjl, 535, L99

\bibitem[Ott et al.(2005)]{ott05} Ott, J., Walter, F., \& 
Brinks, E.\ 2005, \mnras, 358, 1453

\bibitem[Papaderos et al.(2006)]{papaderos06} Papaderos, P., 
Izotov, Y.~I., Guseva, N.~G., Thuan, T.~X., \& Fricke, K.~J.\ 2006, \aap, 
454, 119

\bibitem[Pastoriza et al.(1993)]{pastoriza93} Pastoriza, M.~G., 
Dottori, H.~A., Terlevich, E., Terlevich, R., \& Diaz, A.~I.\ 1993, \mnras, 
260, 177

\bibitem[Paturel et al.(2003)]{paturel03} Paturel, G., Theureau, G.,
Bottinelli, L., Gougenheim, L., Courdreau-Durand, N., Hallet, N., \& Petit,
C.\ 2003, \aap, 412, 57

\bibitem[P\'erez-Gonz\'alez et al.(2006)]{perezgonzalez06}
P\'erez-Gonz\'alez, P. et al.\ 2006, \apj, 648, 987

\bibitem[Pettini \& Pagel(2004)]{pettini04} Pettini, M., \& 
Pagel, B.~E.~J.\ 2004, \mnras, 348, L59

\bibitem[Pilyugin et al.(2004)]{pilyugin04} Pilyugin, L.~S., 
V{\'{\i}}lchez, J.~M., \& Contini, T.\ 2004, \aap, 425, 849

\bibitem[Pilyugin \& Thuan(2005)]{pilyugin05} Pilyugin, L.~S., \& 
Thuan, T.~X.\ 2005, \apj, 631, 231

\bibitem[Pilyugin et al.(2006)]{pilyugin06} Pilyugin, L.~S., 
Thuan, T.~X., \& V{\'{\i}}lchez, J.~M.\ 2006, \mnras, 367, 1139

\bibitem[Popescu et al.(2000)]{popescu00} Popescu, C.~C., 
Misiriotis, A., Kylafis, N.~D., Tuffs, R.~J., \& Fischera, J.\ 2000, \aap, 
362, 138

\bibitem[Pustilnik et al.(2003)]{pustilnik03} Pustilnik, S.~A., 
Kniazev, A.~Y., Pramskij, A.~G., Ugryumov, A.~V., \& Masegosa, J.\ 2003, 
\aap, 409, 917

\bibitem[Pustilnik et al.(2004)]{pustilnik04} Pustilnik, S.~A., 
Pramskij, A.~G., \& Kniazev, A.~Y.\ 2004, \aap, 425, 51

\bibitem[Pustilnik \& Martin(2007)]{pustilnik07} Pustilnik, S.~A., \& Martin,
J.-M.\ 2007, \aap, 464, 859

\bibitem[Raimann et al.(2000)]{raimann00} Raimann, D., 
Storchi-Bergmann, T., Bica, E., Melnick, J., \& Schmitt, H.\ 2000, \mnras, 
316, 559

\bibitem[Reach et al.(2005)]{reach05} Reach, W.~T., et al.\ 
2005, \pasp, 117, 978

\bibitem[Rieke \& Low(1972)]{rieke72} Rieke, G. H., \& Low, F. J. 1972, \apjl, 176, L95

\bibitem[Rieke \& Lebofsky(1979)]{rieke79} Rieke, G.~H., \& 
Lebofsky, M.~J.\ 1979, \araa, 17, 477

\bibitem[Rieke \& Lebofsky(1985)]{rieke85} Rieke, G.~H., \& 
Lebofsky, M.~J.\ 1985, \apj, 288, 618

\bibitem[Rieke et al.(1993)]{rieke93} Rieke, G. H., Loken, K., Rieke, M. J., \& Tamblyn, P. 1993, \apj, 412, 99

\bibitem[Rieke et al.(2004)]{rieke04} Rieke, G.~H., et al.\ 
2004, \apjs, 154, 25

\bibitem[Roche et al.(1991)]{roche91} Roche, P. F., Aitken, D. K., Smith, C. H., \&
Ward, M. J. 1991, \mnras, 248, 606

\bibitem[Roennback \& Bergvall(1995)]{roennback95} Roennback, J., 
\& Bergvall, N.\ 1995, \aap, 302, 353

\bibitem[Rupke et al.(2007)]{rupke07} Rupke, D. S. N., Veilleux, S., \& Baker, A. J. 2007, astro-ph/0708.1766

\bibitem[Sanders \& Mirabel (1996)]{sanders96} Sanders, D. B., \& Mirabel, I. F. 1996, \araa, 34, 749

\bibitem[Sauty et al.(2003)]{sauty03} Sauty, S., et al.\ 2003, 
\aap, 411, 381

\bibitem[Sauvage et al.(2005)]{sauvage05} Sauvage, M., Tuffs, 
R.~J., \& Popescu, C.~C.\ 2005, Space Science Reviews, 119, 313

\bibitem[Schmitt et al.(2006)]{schmitt06} Schmitt, H.~R., 
Calzetti, D., Armus, L., Giavalisco, M., Heckman, T.~M., Kennicutt, R.~C., 
Jr., Leitherer, C., \& Meurer, G.~R.\ 2006, \apjs, 164, 52

\bibitem[Searle \& Sargent(1972)]{searle72} Searle, L., \& 
Sargent, W.~L.~W.\ 1972, \apj, 173, 25

\bibitem[Shaw \& Dufour(1995)]{shaw95} Shaw, R.~A., \& Dufour, 
R.~J.\ 1995, \pasp, 107, 896 

\bibitem[Shi et al.(2005)]{shi05} Shi, F., Kong, X., Li, C., 
\& Cheng, F.~Z.\ 2005, \aap, 437, 849

\bibitem[Shupe et al.(1998)]{shupe98} Shupe, D.~L., Fang, F., 
Hacking, P.~B., \& Huchra, J.~P.\ 1998, \apj, 501, 597

\bibitem[Skillman et al.(1994)]{skillman94} Skillman, E.~D., 
Televich, R.~J., Kennicutt, R.~C., Jr., Garnett, D.~R., \& Terlevich, E.\ 
1994, \apj, 431, 172

\bibitem[Skillman et al.(1989)]{skillman89} Skillman, E.~D., 
Kennicutt, R.~C., \& Hodge, P.~W.\ 1989, \apj, 347, 875

\bibitem[Skillman \& Kennicutt(1993)]{skillman93} Skillman, E.~D., 
\& Kennicutt, R.~C., Jr.\ 1993, \apj, 411, 655

\bibitem[Skillman et al.(1998)]{skillman98} Skillman, E.~D., 
Terlevich, E., \& Terlevich, R.\ 1998, Space Science Reviews, 84, 105

\bibitem[Skrutskie et al.(2006)]{skrutskie06} Skrutskie, M.~F., et
al.\ 2006, \aj, 131, 1163

\bibitem[Smith et al.(2007)]{smith07} Smith, J.-D.~T., et al.\ accepted to
\apj

\bibitem[Soifer et al.(1987)]{soifer87} Soifer, B. T., Neugebauer, G., \& Houck, J. R. 1987,
\araa, 25, 187

\bibitem[Stansberry et al.(2007)]{stansberry07} Stansberry, J.~A., 
et al.\ 2007, \pasp, 119, 1038

\bibitem[Storchi-Bergmann et al.(1994)]{storchi-bergmann94} 
Storchi-Bergmann, T., Calzetti, D., \& Kinney, A.~L.\ 1994, \apj, 429, 572

\bibitem[Storchi-Bergmann et al.(1995)]{storchi-bergmann95} 
Storchi-Bergmann, T., Kinney, A.~L., \& Challis, P.\ 1995, \apjs, 98, 103

\bibitem[Swaters et al.(2002)]{swaters02} Swaters, R.~A., van Albada, T.~S.,
van der Hulst, J.~M., \& Sancisi, R.\ 2002, \aap, 390, 829

\bibitem[Telesco(1988)]{telesco88} Telesco, C.~M.\ 1988, \araa, 
26, 343

\bibitem[Thompson et al.(2006)]{thompson06} Thompson, R.~I., 
Sauvage, M., Kennicutt, R.~C., Jr., Engelbracht, C.~W., \& Vanzi, L.\ 2006, 
\apj, 638, 176

\bibitem[Thuan \& Martin(1981)]{thuan81} Thuan, T.~X., \& Martin, G.~E.\ 1981,
\apj, 247, 823

\bibitem[Thuan et al.(1999)]{thuan99} Thuan, T. X., Sauvage, M., \& Madden, S. 1999, \apj, 516, 783

\bibitem[Thuan et al.(2004)]{thuan04} Thuan, T.~X., Hibbard, J.~E., \&
L(\'e)vrier, F.\ 2004, \aj, 128, 617

\bibitem[Thuan \& Izotov(2005)]{thuan05} Thuan, T.~X., \& 
Izotov, Y.~I.\ 2005, \apjs, 161, 240

\bibitem[Tully et al.(2006)]{tully06} Tully, R.~B., et al.\ 
2006, \aj, 132, 729

\bibitem[Vacca \& Conti(1992)]{vacca92} Vacca, W.~D., \& Conti, 
P.~S.\ 1992, \apj, 401, 543

\bibitem[van Breugel et al.(1985)]{vanbreugel85} van Breugel, W., 
Filippenko, A.~V., Heckman, T., \& Miley, G.\ 1985, \apj, 293, 83

\bibitem[van Zee et al.(1997)]{vanzee97} van Zee, L., Haynes, 
M.~P., \& Salzer, J.~J.\ 1997, \aj, 114, 2479

\bibitem[van Zee et al.(1998)]{vanzee98} van Zee, L., Skillman, E.~D., \&
Salzer, J.~J.\ 1998, \aj, 116, 1186

\bibitem[van Zee(2000)]{vanzee00} van Zee, L.\ 2000, \aj, 119, 
2757

\bibitem[van Zee \& Haynes(2006)]{vanzee06} van Zee, L., \& 
Haynes, M.~P.\ 2006, \apj, 636, 214

\bibitem[Vanzi et al.(1996)]{vanzi96} Vanzi, L., Rieke, G.~H., 
Martin, C.~L., \& Shields, J.~C.\ 1996, \apj, 466, 150

\bibitem[Vanzi et al.(2000)]{vanzi00} Vanzi, L., Hunt, L.~K., 
Thuan, T.~X., \& Izotov, Y.~I.\ 2000, \aap, 363, 493

\bibitem[Werner et al.(2004)]{werner04} Werner, M.~W., et al.\ 
2004, \apjs, 154, 1

\bibitem[Wu et al.(2006)]{wu06} Wu, Y., Charmandaris, V., Hao, L., Brandl, B.
R., Bernard-Salas, J., Spoon, H. W., \& Houck, J. R. 2006, \apj, 639, 157

\bibitem[Young et al.(2003)]{young03} Young, L.~M., van Zee, L., Lo, K.~Y.,
Dohm-Palmer, R.~C., \& Beierle, M.~E.\ 2003, \apj, 592, 111

\bibitem[Zamorano \& Rego(1986)]{zamorano86} Zamorano, J., \& 
Rego, M.\ 1986, \aap, 170, 31

\end{thebibliography}
\end{document}